\documentclass[%
superscriptaddress,
reprint,
nofootinbib,
 amsmath,amssymb,
aps,
prx,
]{revtex4-2}

\usepackage{graphicx}
\usepackage{dcolumn}
\usepackage{bm}
\usepackage[
    colorlinks=true,
    linkcolor=blue,
    urlcolor=blue,
    citecolor=blue]{hyperref}

\usepackage{cancel}

\usepackage[dvipsnames]{xcolor}

\usepackage[arrows]{ezedits}
\defineEdit{DK}{\color[rgb]{0.9,0.1,0.9}}{\color[rgb]{0.9,0.1,0.9}}
\defineEdit{AW}{\color[rgb]{1.0,0.2,0.5}}{\color[rgb]{1,0.2,0.5}}
\defineEdit{DG}{\color[rgb]{.4,.7,0.5}}{\color[rgb]{.4,.7,0.5}}
\defineEdit{TL}{\color[rgb]{0,.2,1}}{\color[rgb]{0,.2,1}}

\begin{document}

\title{Experimental determination of comagnetometer response to non-magnetic spin couplings}
\title{Universal determination of comagnetometer response to spin couplings}
\author{Mikhail Padniuk}
\email{michal.padniuk@doctoral.uj.edu.pl}
\affiliation{Marian Smoluchowski Institute of Physics, Jagiellonian University in Krakow, Łojasiewicza 11, 30-348, Krakow, Poland}

\author{Emmanuel Klinger}
\affiliation{Johannes Gutenberg-Universit\"at Mainz, 55128 Mainz, Germany}
 \affiliation{Helmholtz-Institut Mainz, GSI Helmholtzzentrum f{\"u}r Schwerionenforschung, 55128 Mainz, Germany}
 \affiliation{Universit\'e de Franche-Comt\'e, SupMicroTech-ENSMM, UMR 6174 CNRS, Institut FEMTO-ST, 25000 Besan\c{c}on, France}

\author{Grzegorz Łukasiewicz}
\affiliation{Marian Smoluchowski Institute of Physics, Jagiellonian University in Krakow, Łojasiewicza 11, 30-348, Krakow, Poland}

\author{Daniel Gavilan-Martin}
\affiliation{Johannes Gutenberg-Universit\"at Mainz, 55128 Mainz, Germany}
\affiliation{Helmholtz-Institut Mainz, GSI Helmholtzzentrum f{\"u}r Schwerionenforschung, 55128 Mainz, Germany}

\author{Tianhao Liu}
\email{Tragically deceased on July 22nd 2023}
\affiliation{Johannes Gutenberg-Universit\"at Mainz, 55128 Mainz, Germany}
\affiliation{Helmholtz-Institut Mainz, GSI Helmholtzzentrum f{\"u}r Schwerionenforschung, 55128 Mainz, Germany}
\affiliation{Physikalisch-Technische Bundesanstalt Berlin, 13585 Berlin, Germany}

\author{Szymon Pustelny}
\affiliation{Marian Smoluchowski Institute of Physics, Jagiellonian University in Krakow, Łojasiewicza 11, 30-348, Krakow, Poland}

\author{Derek F. Jackson Kimball}
\affiliation{Department of Physics, California State University – East Bay, Hayward, CA 94542, USA}

\author{Dmitry Budker}
\affiliation{Johannes Gutenberg-Universit\"at Mainz, 55128 Mainz, Germany}
\affiliation{Helmholtz-Institut Mainz, GSI Helmholtzzentrum f{\"u}r Schwerionenforschung, 55128 Mainz, Germany}
\affiliation{Department of Physics, University of California, Berkeley, CA 94720, USA}
 
\author{Arne Wickenbrock}
\affiliation{Johannes Gutenberg-Universit\"at Mainz, 55128 Mainz, Germany}
\affiliation{Helmholtz-Institut Mainz, GSI Helmholtzzentrum f{\"u}r Schwerionenforschung, 55128 Mainz, Germany}

\begin{abstract}
We propose and demonstrate a general method to calibrate the frequency-dependent response of self-compensating noble-gas-alkali-metal comagnetometers to arbitrary spin perturbations. This includes magnetic and non-magnetic perturbations like rotations and exotic spin interactions. The method is based on a fit of the magnetic field response to an analytical model. 
The frequency-dependent response of the comagnetometer to arbitrary spin perturbations can be inferred using the fit parameters. We demonstrate the effectiveness of this method by comparing the inferred rotation response to an experimental measurement of the rotation response.
Our results show that experiments relying on zero-frequency calibration of the comagnetometer response can over- or underestimate the comagnetometer sensitivity by orders of magnitude over a wide frequency range. 
Moreover, this discrepancy accumulates over time as operational parameters tend to drift during comagnetometer operation. The demonstrated calibration protocol enables accurate prediction and control of comagnetometer sensitivity to, for example, ultralight bosonic dark-matter fields coupling to electron or nuclear spins, as well as accurate monitoring and control of the relevant system parameters.

\end{abstract}

\maketitle


\section{Introduction}
Over two decades of development, self-compensating noble-gas-alkali-metal comagnetometers have been used for fundamental physics tests \cite{vasilakis2009limits,terrano2021comagnetometer,blocholdcomag,wei2022ultrasensitive} and precise rotation measurements with potential applications for navigation in challenging conditions \cite{kornack2005nuclear,jiang2018parametrically,Liu2022Comag,Liang2022Biaxial,wei2022ultrasensitive}. Recently, these systems have gained attention as promising tools to realize long-lived quantum memories \cite{katz2022quantum,shaham2022strong,katz2022optical}. In this case, the potential arises from the long coherence times of the noble gas spins and efficient access to the noble gas via the alkali species. 

In a comagnetometer operating under the right working conditions, called the self-compensating regime, the nuclear magnetization adiabatically cancels transverse \footnote{Conventionally, the field oriented along the pump beam propagation direction is called the longitudinal field and fields orthogonal to the pump beam propagation direction are called transverse fields.} external magnetic fields experienced by the alkali-metal spins. These conditions are achieved at the so-called compensation point, at which the externally applied longitudinal magnetic field approximately cancels the total field experienced by the electron spins of the alkali-metal atoms [the precise definition of the compensation point is provided in Eq.\,\eqref{eq:compensation_point_definition}]. It is noteworthy that the total field includes not only the external leading magnetic field, but also the effective field arising from rotations, exotic fields coupling to spins, and collisional interaction of alkali and noble-gas atoms \cite{walker1997spin}. 

In turn, at the compensation point, even though the alkali-metal spins constitute a highly sensitive zero-field magnetometer, the comagnetometer signal is insensitive to drifts and fluctuations of external transverse magnetic fields at frequencies below the noble-gas Larmor frequency. 

As highly-sensitive atomic magnetometers are often limited by the stability of the magnetic environment, compensation of the slowly drifting fields makes the comagnetometers an attractive choice for applications that require continuous measurement for long time periods (e.g., hours or even days). Examples of such measurements are dark-matter searches \cite{kimball2022search,terrano2021comagnetometer} and measurements of electric dipole moments (EDMs) of fundamental particles \cite{chupp2019electric}, as well as, as mentioned above, navigation applications \cite{fang2012advances,cai2018error}. 
To perform such applications, however, the frequency response of the comagnetometers to the non-magnetic perturbations (e.g., rotations) has to be accurately known, which requires a reliable method of calibration.

In the search for exotic spin couplings, presently, comagnetometer-based experiments provide some of the most stringent limits on Lorentz invariance violation \cite{Smiciklas2011}, spin-dependent gravitational interactions \cite{sheng2023spingravity} and spin-dependent fifth forces \cite{Lee2018}. Many proposed extensions of the Standard Model (SM) predict the existence of new ultralight bosons \cite{Gra15,co2020predictions,Svr06,Arv10}, which could explain dark matter. Such ultralight bosonic dark matter could interact with SM particles over a variety of portals \cite{graham2016dark,safronova2018search,kimball2022search}, leading to oscillations of fundamental constants and nuclear and electronic EDMs, as well as torques on spins. The mass of these ultralight bosons could be anywhere between 10$^{-22}$ and 10\,eV, which results in a large boson--mass/coupling--strength parameter space to be explored. To date, several searches of such interactions have been published and more are on the way \cite{graham2015experimental,graham2018spin,kimball2022search,afach_what_2023}.

For gradient-coupled axion-like particle dark matter \cite{kimball2022search}, self-compensating comagnetometers place the most stringent limits in the frequency range from 0.01 to 10\,Hz, corresponding to a mass range between $4 \times 10^{-17}$ and $4 \times 10^{-14}$\,eV \cite{Lee2023,wei_dark_2023} (overall these are the most stringent limits in any mass range). 
Other experiments are looking even beyond this mass range \cite{bloch2022New,Bloch:2023:natComm,centers2021stochastic}. In order to characterize a signal due to exotic interactions or place meaningful bounds, understanding the frequency response of the system to exotic-physics-related fields is of utmost importance.

Self-compensating comagnetometers also form the core of the upgraded advanced version of the Global Network of Optical Magnetometers for Exotic physics searches (GNOME) \cite{pustelny_global_2013,afach_characterization_2018,afach_search_2021,afach_what_2023}. GNOME is an international network of spin-state sensors \cite{kimball2023probing} (e.g., atomic magnetometers \cite{budker_optical_2007,budker_optical_2013}), currently with 14 stations, looking for spatio-temporally correlated signatures of ultralight bosonic dark matter. The sought-after signals could be generated by compact composite exotic physics objects such as axion-like field domain walls \cite{Pos13,Mas20,afach_search_2021,kim2022machine}, axion stars \cite{Kim18AxionStars},
gravitationally bound axion halos \cite{banerjee2020relaxion,banerjee2020searching, budker2023generic}, 
but also bursts of ultralight bosons emitted by cataclysmic astrophysical events (such as binary black hole mergers \cite{dailey2021quantum}) and stochastic fluctuations of the ultralight fields \cite{centers2021stochastic,masia2023intensity}. The upgrade in the network is driven by the improved sensitivity of self-compensating comagnetometers to nuclear spin couplings, which is three (proton) and six (neutron) orders of magnitude better than that of previously used alkali-vapor-only magnetometers. Here the frequency response is crucial to understand how time-dependent signals measured with the comagnetometer differ from the predicted transient exotic field signals.
 
Previously, we studied the frequency response of self-compensating comagnetometers numerically \cite{padniuk2022response}. However, a direct measurement of the frequency response to exotic fields is so far impossible, since the fields have not been observed. 
 
In this work, we propose and demonstrate a calibration method to infer the frequency response of a self-compensating comagnetometer to general spin perturbations. The proposed calibration method involves an easy-to-implement protocol to measure the magnetic field frequency response, fitting it with an analytical model, and subsequently using the fit parameters to deduce the response to non-magnetic couplings. To validate the method, we built a self-compensating comagnetometer on a rotation stage and directly measured the frequency response to rotations. The experimentally measured results show excellent agreement with the model. Additionally, we discuss how the frequency response changes as a function of the applied leading magnetic field. We experimentally show how errors in the assumed field dramatically affect the interpretation of measurement results. In general, the method enables comagnetometer-based searches for new physics and accurate rotation sensing over a broad frequency range.

The paper is structured as follows: first we briefly lay out the theoretical framework, relate the magnetic frequency response to the frequency response of other perturbations and explain how to measure the magnetic frequency response (Sec.\,\ref{sec:theory}). Then we describe experimental setup and procedure (Sec.\,\ref{sec:ExpSetup}) before presenting and discussing the obtained results (Sec.\,\ref{sec:results}). Finally, we conclude in Sec.\,\ref{sec:Conclusions}.

\section{Theory of comagnetometer frequency response} \label{sec:theory}
 The dynamics of two overlapping spin ensembles have been actively studied since 2002 \cite{kornack2002dynamics}. 
Although theoretical and experimental considerations regarding the frequency response of the self-compensating comagnetometer system to a magnetic field have been published in multiple references \cite{kornack_test_nodate,vasilakis_precision_nodate, FanIEEE2019}, the theoretical considerations of the frequency response to exotic spin couplings have been published in Ref.\,\cite{kornack_nuclear_2005, padniuk_response_2022}. 

Here, we review the main results from Ref.~\cite{padniuk2022response} and its supplemental material and utilize them to construct the frequency response to rotations and exotic fields based on the magnetic field frequency response.

The coupled evolution of the alkali-metal polarization $\mathbf{P^e}$ and noble-gas polarization $\mathbf{P^n}$ can be described with a coupled system of two Bloch equations (also known as Bloch-Hasegawa equations) \cite{Hasegawa1959, kornack_dynamics_2002} 
\begin{widetext}
\begin{equation}
\begin{split}
           \frac{d \mathbf{P^e}}{d t} =& \frac{1}{q}\gamma_e \mathbf{(B+\alpha_e b}+\lambda M^n\mathbf{ P^n)\times P^e}
                +\frac{1}{q}[R_\text{se}^{en}(\mathbf{P^n-P^e}) + ({\mathbf{P}}^{\mathbf{e}}_0-{\mathbf{P^e}})R^e ], \\
            \frac{d \mathbf{P^n}}{d t} =&
            \gamma_n \mathbf{(B+\alpha_n b}+\lambda M^e\mathbf{ P^e)\times P^n}+ R_\text{se}^{ne}(\mathbf{P^e-P^n})- R^n\mathbf{P^n}.
\label{eq:BHE_1}
\end{split}
\end{equation}

\end{widetext}
Here, $\gamma_e$ and $\gamma_n$ are the gyromagnetic ratios of the free electron spin and the noble-gas nuclear spin, respectively.  The slowing-down factor of the alkali spins $q$, reduces the electronic gyromagnetic ratio $\gamma_e$ due to the hyperfine interaction and redistribution of atoms over the hyperfine levels due to spin-exchange collisions \cite{Allred2002}. The collisional coupling constant $\lambda = 2\kappa_0\mu_0/3$ is characteristic for a given mixture of noble gas and alkali-metal vapor and is defined with vacuum permeability $\mu_0$ and spin-exchange enhancement factor $\kappa_0$. The latter results from the overlap of the alkali electron wave function and the nucleus of the noble gas \cite{baranga1998}. $M^e$ represents the electron magnetization, while $M^n$ represents the nuclear magnetization of the fully polarized spin species. The rates of polarization transfer from the electronic to the nuclear species (and vice versa) by spin-exchange collisions are denoted by $R^{en}_{se}$ ($R^{ne}_{se}$). $R^e$ represents the relaxation rate of the alkali-metal polarization due to all relaxation processes, including these related to the pump light. The equilibrium electronic polarization, $\mathbf{P}^{\mathbf{e}}_0$, results from optical pumping by the pump light. $R^n$ is the relaxation rate of the nuclear polarization of the noble gas.  Generic (i.e., magnetic or nonmagnetic) external perturbations are introduced by $\mathbf{B}$ and $\mathbf{b}$. The constant external magnetic field $\mathbf{B}$ sets the operation conditions (i.e., the self-compensation mode). The vector $\mathbf{b}$ represents possible perturbations with coupling constants $\alpha_e$ and $\alpha_n$, where the subscripts denote coupling to electron and nuclear polarization, respectively. The coupling constants $\alpha_e$ and $\alpha_n$ differ for magnetic fields, rotations, and exotic couplings and are given in Table\,\ref{tab:coupling_parameters}.

\begin{table}[tb]
 \def\arraystretch{2}
\begin{tabular}{lccc}
     \toprule
     Coupling type & $\mathbf{b}$&$\alpha_e$ & $\alpha_n$  \\ \hline
     Magnetic field & $\mathbf{B_{ext}}$ & 1 &1 \\
     Rotation & $\frac{\bm{\Omega}}{\gamma_n}$& $q\frac{\gamma_n}{\gamma_e}$ & $1$ \\
     Exotic  neutron coupling &$\frac{\hbar\sigma^n_{neu}\chi_{neu}}{\gamma_n}\mathbf{\Xi_{neu}}$&  $(q-1)\frac{\gamma_n}{\gamma_e}\frac{\sigma^e_{neu}}{\sigma^n_{neu}}$ &1\\ 
        Exotic proton coupling &$\frac{\hbar\sigma^n_{p}\chi_{p}}{\gamma_n}\mathbf{\Xi_{p}}$& $(q-1)\frac{\gamma_n}{\gamma_e}\frac{\sigma^e_{p}}{\sigma^n_{p}}$&1\\ 
     Exotic electron coupling & $\frac{\hbar\chi_{e}}{\gamma_n}\bm{\Xi_e}$ &$q\frac{\gamma_n}{\gamma_e}$& 0 \\
     \toprule
\end{tabular}
\caption{Coupling constants $\alpha_e$ and $\alpha_n$ used in Eq.~\eqref{eq:BHE_1} to parameterize spin couplings of different origin. The notation used in the table is the following: $\mathbf{B_{ext}}$ corresponds to an additional external magnetic field transverse to external magnetic field $\mathbf{B}$, $\mathbf{\Omega}$ represents the angular velocity vector of a mechanical rotation of the setup, $\bm{\Xi_{i}}$ denotes an exotic perturbation and $\chi_{i}$ characterizes the coupling strength with the subscript $i$ being ``$neu$'' for neutrons, $p$ for protons, and $e$ for electrons. Exotic couplings to nucleons also affect the electronic polarisation via the hyperfine interaction. This is considered with constants $\sigma^{e}_{neu}$ and $\sigma^{e}_{p}$, for exotic neutron and proton couplings, respectively.
$\sigma^{n}_{\text{neu}}$($\sigma^{n}_{p}$) is the neutron (proton) fraction of the noble-gas nucleon spin, and $\sigma^{e}_{neu}$($\sigma^{e}_{p}$) is the neutron (proton) fraction of the nuclear spin of the alkali metal \cite{kimball_nuclear_2015}. }
\label{tab:coupling_parameters}
\end{table}

We are interested in the frequency response of the system to generic spin perturbations and an experimental way to test this. For the first task, we derive an analytical expression for the frequency response, assuming the system operates near the self-compensating point. The chosen value of $\mathbf{B}$ is determined by the equilibrium electronic spin polarization along the pump-beam axis, i.e., in the longitudinal direction. Since the measured signal is determined by the transverse polarization, we restrict our analysis to the response of the system to transverse fields. 

The comagnetometer is tuned to the self-compensating regime by setting the constant external magnetic field $\mathbf{B}$ to the compensation point $\mathbf{B_c}$ \cite{kornack_dynamics_2002} 
\begin{equation}
    \mathbf{B_c} = -\lambda( M^n\mathbf{P}_0^n + M^e \mathbf{P}_0^e)\,,
\label{eq:compensation_point_definition}
\end{equation}
 where $\mathbf{P}_0^n = R_{se}^{ne}\mathbf{P_0^e}/(R^n+R_{se}^{ne})$ is the equilibrium nuclear polarization. We are interested in the operation of the system around the compensation point and introduce the field difference (detuning) $\bm{\Delta_B}$ relative to the compensation point
\begin{equation}
\label{eq:detunning_definition}
    \bm{\Delta_B} = \mathbf{B_c-B}\,.
\end{equation}
Hereafter, we replace the vector quantity $\bm{\Delta_B}$ by its z-component $\Delta_{B_z}$, assuming that the other parts of the field are zero. Furthermore, we separate $\mathbf{P}_0^n$ and $\mathbf{P}_0^e$ into longitudinal ($P^n_\parallel$, $P^e_\parallel$) and transverse ($P^n_\perp$, $P^e_\perp$) components. Close to the compensation point, the effective magnetic field experienced by each spin species from the polarization of the other species and the applied compensating fields are much larger ($\sim$ nT) than the considered external spin perturbations ($\sim$ fT--pT). We utilize the small-angle approximation, assuming that the longitudinal polarizations to be constant and equal to their equilibrium values. This allows us to linearize the coupled Bloch equations by separating the constant longitudinal part and the time-varying transverse part. This approximation, along with Eq.~\eqref{eq:compensation_point_definition}, result in the following form of the Bloch equations 
\begin{widetext}    
\begin{equation}
\label{eq:compact_BHE}
\begin{split}
        \frac{d{P^e_\perp}}{dt} &= -i \frac{\gamma_e}{q}\bigg[\big(\alpha_e b_\perp+\lambda M^n{P^n_\perp}\big) {P^e_\parallel}
        -{\big(\Delta_{B_z}}+\lambda M^e{P^e_\parallel}\big) P^e_\perp
        \bigg]
    -\frac{R_e}{q} P^e_\perp \\
        \frac{d{P^n_\perp}}
        {dt}&=
        -i\gamma_n\bigg[\big(\alpha_n {b}_\perp+\lambda M^e {P^e_\perp}\big) {P^n_\parallel}-\big(\Delta_{B_z}+\lambda M^n {P^n_\parallel}\big){P^n_\perp}\bigg] 
        -R_nP^n_\perp
\end{split}
,
\end{equation}
\end{widetext}
where all vector quantities are separated into the real longitudinal ($\parallel$) and complex transverse ($\perp$) parts in a similar manner to polarisation \footnote{For any vector $\mathbf{x}$ in the Cartesian coordinate system, $x_\parallel = x_z$ and $x_\perp = x_x+i x_y$, where we assume that the polarization axis is co-linear with $\mathbf{z}$.}. The total relaxation rates $R_e$ and $R_n$ take into account all relaxation processes, including polarization transfer due to spin exchange.

In this work, we are interested in the frequency response of the comagnetometer to a generic spin perturbation. This can be obtained by solving the Bloch equations \eqref{eq:compact_BHE} with oscillating perturbation of the amplitude $b_0$ at frequency $\omega$
\begin{equation}
    b = i b_0 \sin(\omega t)\,.
    \label{eq:sine}
\end{equation}
From the analysis of Eq.\,\eqref{eq:compact_BHE}, one obtains that for such a perturbation the transverse polarization oscillates at the same frequency and hence can be written as
\begin{equation}
    P_\perp^e(t) = \frac{1}{2}\bigg(P^{e+}_\perp(\omega) e^{i\omega t}- {P^{e-}_\perp(\omega)}e^{-i\omega t}\bigg)\,,
\label{eq:time_solution_complex_polarisation}
\end{equation}
where the amplitudes of the transverse electronic polarizations $P^{e\pm}_\perp$ are given by
\begin{widetext}
    \begin{equation}
         P_\perp^{e\pm}(\omega) = -\frac{\gamma_eP^e_\parallel b_0}{q} \frac{\omega_n(\alpha_e-\alpha_n)+(\pm\omega+\gamma_n\Delta_{B_z}-i R_n)\alpha_e}{(\pm\omega+\omega_e+\gamma_e\Delta_{B_z}/q-iR_e/q)(\pm \omega +\omega_n-+\gamma_n \Delta_{B_z}-iR_n)-\omega_e\omega_n}\,.
    \label{eq:time_solution_complex_polarisation_detail}
    \end{equation}
\end{widetext}
In Eq.\,\eqref{eq:time_solution_complex_polarisation_detail},  $\omega_e = \gamma_e\lambda M^e P^e_\parallel/q$ and $\omega_n = \gamma_n\lambda M^nP^n_\parallel$ are the Larmor frequencies of the electron and nuclear polarizations at the compensation point. Dividing the transverse electronic polarization amplitudes by the applied perturbation amplitude results in the frequency response:
 \begin{equation}
\label{eq:P_x_spectrum}
   \mathcal{F}(\omega)=\frac{P^{e}_\perp (\omega)}{b_0}\,. 
\end{equation}
This gives the frequency response to the perturbation at a single frequency $\omega$. A procedure for complete determination of $\mathcal{F}(\omega)$ at all frequencies via measurement and fitting of a comagnetometer's response to controlled transverse magnetic field perturbations is the key result of this work. 

\subsection{Measurement of the frequency response with a pulse}
\label{sec:measurement_of_freq_response_with_pulse}
The response of the self-compensating comagnetometer is linear in small external perturbations with respect to the compensating field; therefore, to obtain the spectrum of the response $P_x^e(\omega)$, the frequency response of the system $\mathcal{F}(\omega)$ is multiplied by the specific perturbation spectrum $b(\omega)$:
\begin{equation}
\label{eq:P_x_spectrum}
    P_x^e(\omega) = \sqrt{2\pi}\mathcal{F}(\omega)b(\omega)\,. 
\end{equation}

Therefore, the frequency response allows a quantitative calibration of the system and its parameters and can be measured with a known perturbation. 

One way to measure the frequency response is to excite all possible frequencies by applying a step change of the considered perturbation (magnetic field or rotation). A step change in the time domain is described by the Heaviside theta function $\Theta(t)$ and step amplitude $b_0$
\begin{equation}
\label{eq:test_perturbation}
    b_{test}(t) = b_0\Theta(-t)\,,
\end{equation}
which has the following spectrum
\begin{equation}
\label{eq:theta_heaviside_spectrum}
    b_{test}(\omega)= \mathfrak{F}[b_{test}(t)] = \frac{1}{\sqrt{2\pi}}\frac{b_0}{i\omega} +b_0\sqrt{\frac{\pi}{2}}\delta(\omega).
\end{equation}
Applying this spectrally wide perturbation allows us to determine the complete frequency response in a single measurement. 
The spectral amplitude of the Heaviside theta function changes with frequency, which has to be taken into account to get to the frequency response. We do this by performing a numerical time derivative of the response data. In Fourier space, this is equivalent to a multiplication with $\omega$ and appears simpler. Measuring the response to a perturbation, performing a numerical time derivative of the data and applying a Fourier transform to the result yields the frequency response of the comagnetometer projected onto the measurement axis:
\begin{equation}
\label{eq:derevative_spectrum}
    \mathfrak{F}\bigg[\frac{dP^e_x(t)}{dt}\bigg]= i\omega P^e_x(\omega)= b_0 \mathcal{F}(\omega)\,,
\end{equation}
where we took into account Eqs.~\eqref{eq:P_x_spectrum}, \eqref{eq:test_perturbation} and \eqref{eq:theta_heaviside_spectrum}. Hence, the frequency response of the system can be determined by calculating the Fourier transform of the (numerical) time derivative of the data obtained from the system response to a step change in the amplitude of the considered perturbation 
\begin{equation}
\label{eq:expression_for_experimental_freq_resp}
    \mathcal{F}(\omega) = \frac{1}{b_0}\mathfrak{F}\bigg[\frac{dP^e_x(t)}{dt}\bigg]\,.
\end{equation}

\subsection{From magnetic-field to the total response of the comagnetometer}
Equation~\eqref{eq:time_solution_complex_polarisation_detail} shows that the response to magnetic and non-magnetic perturbations is described with the same set of parameters. When these parameters are known, the system response to perturbations of arbitrary nature and directions can be constructed. Furthermore, the parameters are accessible by measuring any specific (e.g., magnetic) frequency response and fitting it with the appropriate model. This way, the operating regime and the generic frequency response can be determined solely by measuring the magnetic frequency response.

The magnetic field response [Eq.~\eqref{eq:expression_for_experimental_freq_resp}] is fitted with a model obtained from Eqs.~\eqref{eq:time_solution_complex_polarisation} and ~\eqref{eq:time_solution_complex_polarisation_detail}.
Identifying the real and imaginary parts of the response with the in- and out-of-phase components allows one to define a complex signal,
\begin{equation}
\label{eq:complex_fit_model}
    \mathcal{F}^m = (\mathcal{F}^m_{in}+i\mathcal{F}^m_{out})e^{i\theta}= i(\mathcal{F}^{m}_+-\mathcal{F}^{m*}_-)e^{i\theta}\,, 
\end{equation}
where the star operator denotes the complex conjugate and $\theta$ is a fitting parameter taking into account possible phase shifts. The fitting model obtained from Eq.\,\eqref{eq:time_solution_complex_polarisation_detail} for the response to magnetic fields gives the following relation for the components:
\begin{widetext}
\begin{equation}\label{eq:rotation_field_fitting_model}
    \mathcal{F}_\pm^m(\omega) = -{\color{blue}a}
    \frac{\pm \omega +{\color{Green}\gamma_n }
    {\color{blue}\Delta_{B_z}}-i {\color{blue}{|R_n|}}}
    {(\pm\omega+{\color{blue}\omega_e}+ {\color{blue} \Delta_{B_z}}{\color{Green}\gamma_e/q} -{\color{blue}{|R_e|}})
    (\pm\omega + {\color{blue}\omega_n} +{\color{Green}\gamma_n} {\color{blue}\Delta_{B_z}}-i{{\color{blue}|R_n|}})
    -{\color{blue}\omega_e\omega_n}}\,,    
\end{equation}
\end{widetext}
where fitting parameters are marked {\color{blue} blue} and predefined parameters are marked {\color{Green} green}. The predefined parameters are: $\gamma_n$ -- the gyromagnetic ratio of the noble-gas nuclear spin, $\gamma_e$ -- the electron gyromagnetic ratio, and $q$ -- the slowing-down factor approximated by a constant estimated prior to the calibration. The parameters used for fitting the magnetic response can then be used to construct the response to any other perturbation along both transverse axes according to the following model
\begin{align}
\label{eq:reconstruction_primary_axis}
 \mathcal{F}^{r}_{pr}(\omega) & = 
 (\mathcal{F}^{r}_+-\mathcal{F}^{r*}_-)e^{i(\theta+\pi/2)},\\
 \label{eq:reconstruction_secondary_axis} \mathcal{F}^{r}_{sec} (\omega)&=  (\mathcal{F}^{r}_++\mathcal{F}^{r*}_-)e^{i(\theta+\pi)} ,
\end{align}
where $\mathcal{F}^{r}_{pr}(\omega)$ is the response to fields applied along the primary sensitivity axis, i.e., parallel to the propagation direction of the probe beam, and $\mathcal{F}^{r}_{sec}$ is the response to fields along the secondary sensitivity axis, i.e., orthogonal to pump and probe beam. $\mathcal{F}_\pm^r$ has the following form obtained from Eq.\,\eqref{eq:time_solution_complex_polarisation_detail}
\begin{widetext}
\begin{equation}
    \mathcal{F}^{r}_\pm = -{\color{blue}a}
    \frac{{\color{blue}\omega_n}{\color{magenta}(\alpha_e-\alpha_n)}+(\pm \omega +{\color{Green}\gamma_n }{\color{blue}\Delta_{B_z}}-i |{\color{blue}{R_n}}|){\color{magenta}\alpha_e}}
    {(\pm\omega+{\color{blue}\omega_e}+  {\color{blue} \Delta_{B_z}}{\color{Green}\gamma_e/q}-|{\color{blue}{R_e}}|)
    (\pm\omega + {\color{blue}\omega_n} +{\color{Green}\gamma_n} {\color{blue}\Delta_{B_z}}-i|{{\color{blue}R_n}}|)
    -{\color{blue}\omega_e\omega_n}}\,,
    \label{eq:reconstruction_rotarion_fields}
\end{equation}
\end{widetext}
where the coupling strength parameters marked in {\color{magenta} magenta} are set according to Table~\ref{tab:coupling_parameters}.

\section{Experimental setup}\label{sec:ExpSetup}
\begin{figure}[htb]
    \centering
    \includegraphics[width=0.4\textwidth]{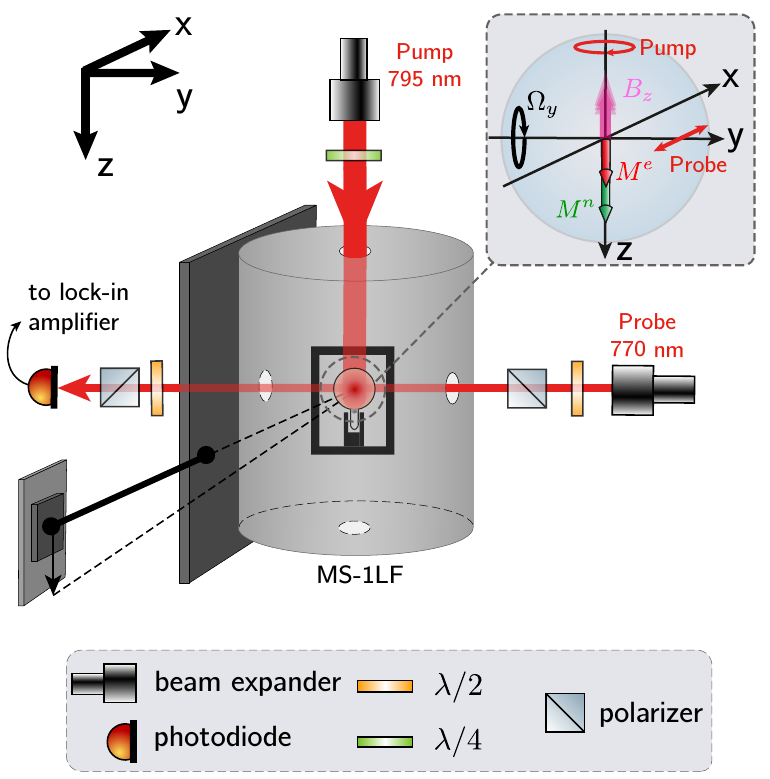}
    \caption{Sketch of the experimental setup. A vapor cell containing $^3$He, N$_2$, $^{87}$Rb and K is situated inside a Tw\^inleaf MS-1LF magnetic shield. The Rb atoms are optically pumped with circularly polarized light from a Toptica TA Pro laser on-resonant to the D1 transition. K and He spins are pumped by spin-exchange collisions with Rb. The readout is realized by measuring the polarization rotation of a probe beam detuned from the K D$_1$ line with a photodiode. The leading field is sinusoidally modulated at 800\,Hz to up-convert the magnetic signal and suppress low-frequency noise. The polarization signal is then demodulated with a lock-in amplifier. The inset shows the directions and polarizations
of the laser beams, magnetic field modulations and the directions of the generated electronic ($M^e$) and nuclear magnetization ($M^n$).}
    \label{fig:setup}
\end{figure}
The experimental setup is shown in Fig.\,\ref{fig:setup} and it is based on the setup reported in Ref.\,\cite{klingerPRA2023}. Briefly, a 20mm diameter spherical vapor cell filled with 3\,amg of $^3$He and 50\,torr of N$_2$ is loaded with a drop of an alkali-metal mixture with 1\%\,$^{87}$Rb and 99\% natural-abundance K (molar fractions). The cell is placed in an oven and heated to $185\,^\circ$ C with an AC resistive heater. The oven assembly is mounted in a Tw\^inleaf MS-1LF magnetic shield. The Rb atoms are optically pumped with about 70\,mW (intensity of about 16\,mW/cm$^2$) circularly polarized light tuned to the center of the rubidium D$_1$ line. Potassium (and helium) atoms are then polarized by spin-exchange collisions with the Rb atoms. The hybrid pumping technique reduces inhomogeneities in the K polarization and improves the efficiency of spin-exchange pumping of the noble-gas atoms \cite{babcock_hybrid_2003, Lee:21}.

The comagnetometer readout is realized by monitoring the polarization rotation of a linearly polarized probe beam detuned about 0.5\,nm towards the shorter wavelength of the potassium D$_1$ line. Low-noise detection of the response to perturbations along the $y$-axis is achieved using lock-in detection and parametric modulation of the $B_z$ field with a sine wave (800\,Hz, 35\,nT peak-to-peak) \cite{zhimin2006parametric,jiang2018parametrically}. The polarization-rotation signal is demodulated with a lock-in amplifier (Zurich Instruments HF2LI) using the first and second harmonics of the modulation frequency. The first-harmonic signal features a linear relationship to the $y$-component of potassium polarization while demodulation at the second harmonic corresponds to measurements of potassium polarization along the $x$-axis, see, e.g., Appendix A in Ref.\,\cite{klingerPRA2023}. 

The comagnetometer is operated around the self-compensation point with an equilibrium compensation field of about 120\,nT, achieved after optimization of the nuclear polarization as discussed in Ref.\,\cite{klingerPRA2023}. At this level of the compensation field, the decay rate of nuclear spins reached 20\,s$^{-1}$. 

Our comagnetometer setup can be rotated about the $y$-axis to controllably generate non-magnetic spin perturbations experimentally. To do so, the shield with the comagnetometer is mounted on a breadboard attached to a slewing bearing (iglidur\textsuperscript \textregistered PRT-01). As shown in Fig.\,\ref{fig:setup}, the system is rotated by actuating a 70.5(1)-cm-long arm with a linear translation stage (Thorlabs MTS50-Z8). Because in the experiment the laser beams are transmitted through fibers and the expanders are attached to the shield, the alignment does not change during rotation (see the inset in Fig.\,\ref{fig:setup}). With this configuration, the applied rotation rate can be approximated by $   \Omega_y(t)\approx v(t)/L$,
where $v(t)$ is the velocity of the translation stage and $L$ is the arm length. This is valid as long as the rotation angle (hence the translation-stage travel) is small enough. As the travel range of our translation stage is limited to 50\,mm, the maximum angle applied to the system is about 64.61(9)\,mrad. With a maximum velocity of 2.4(2)\,mm/s, the highest achievable rotation rate is 3.4(3)\,mrad/s. 

\section{Results and Discussion}\label{sec:results}
The frequency response to magnetic field and rotation was determined based on the response of the comagnetometer to a step perturbation, using the procedure described in Sec.~\ref{sec:measurement_of_freq_response_with_pulse}. 
For magnetic fields, the response was obtained by applying an $80$-pT square pulse along the $y$-axis. The pulse duration was chosen to be sufficiently long (4\,s) to reach the steady state regime before applying the next field value $B_y$. The response to the falling edge of the pulse was utilized to determine the frequency response.  
The same routine was followed to determine the rotation response. However, in this case, the square pulse consisted of three steps: angular acceleration of the rotation stage to a constant rotation rate, sufficiently long (4\,s) constant rotation to reach a steady state, followed by a sudden stop of the motion. The motor driving the rotation stage was accelerated to a speed of 2.0(2)\,mm/s, corresponding to a rotation rate of  $\Omega_y = 2.8(3)$\,mrad/s, which translates to $\approx 14(1)$~pT of effective pseudomagnetic field for the noble-gas spins. 

\begin{figure*}[htb]
    \centering
    \includegraphics[scale = 1]{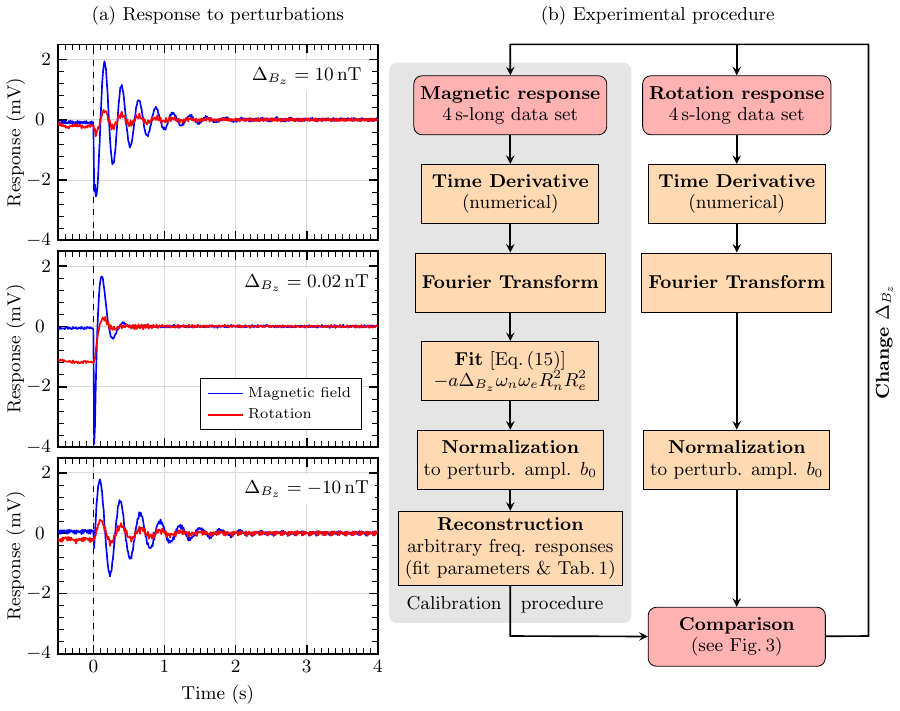}
    \caption{(a) Time series of the comagnetometer response to step changes in rotation and magnetic field under different detunings from the compensation point: below the compensation point (top), at the compensation point (middle), and above the compensation point (bottom). 
     (b) A flowchart illustrating the experimental procedure used to study the response of the comagnetometer to magnetic fields and rotations, for different detunings from the compensation point. Note that the acquisition of the magnetic and rotation responses is separated (in time) by about 8\,s. The gray-shaded area highlights the necessary steps of the calibration procedure proposed and described in this work, see text. 
    }
    \label{fig:temporal_response}
\end{figure*}

A single dataset contains the response to magnetic and rotation pulses. In total 71 datasets were collected for different magnetic detunings from the compensation point [Eq.~\eqref{eq:detunning_definition}], ranging from $-10$ to $15$~nT. A graphical summary of the measurement sequence is presented with a flow chart in Fig.\,\ref{fig:temporal_response}(b).
Figure\,\ref{fig:temporal_response}(a) shows an example time series of the response to magnetic and rotation perturbations for three different leading field conditions: below (upper plot), close to (middle plot), and above the compensation field (lower plot).
In the results presented, one can note the characteristic spin dynamics for the self-compensating regime shown in the middle plot of Fig.\,\ref{fig:temporal_response}(a). It manifests as a strong damping of the magnetic field response, along with the highest response to low-frequency rotation. This can be contrasted with the dynamics away from the compensation point as shown in the upper and lower plots of Fig.\,\ref{fig:temporal_response}(a).

\begin{figure*}
    \centering
    \includegraphics [width = \linewidth]{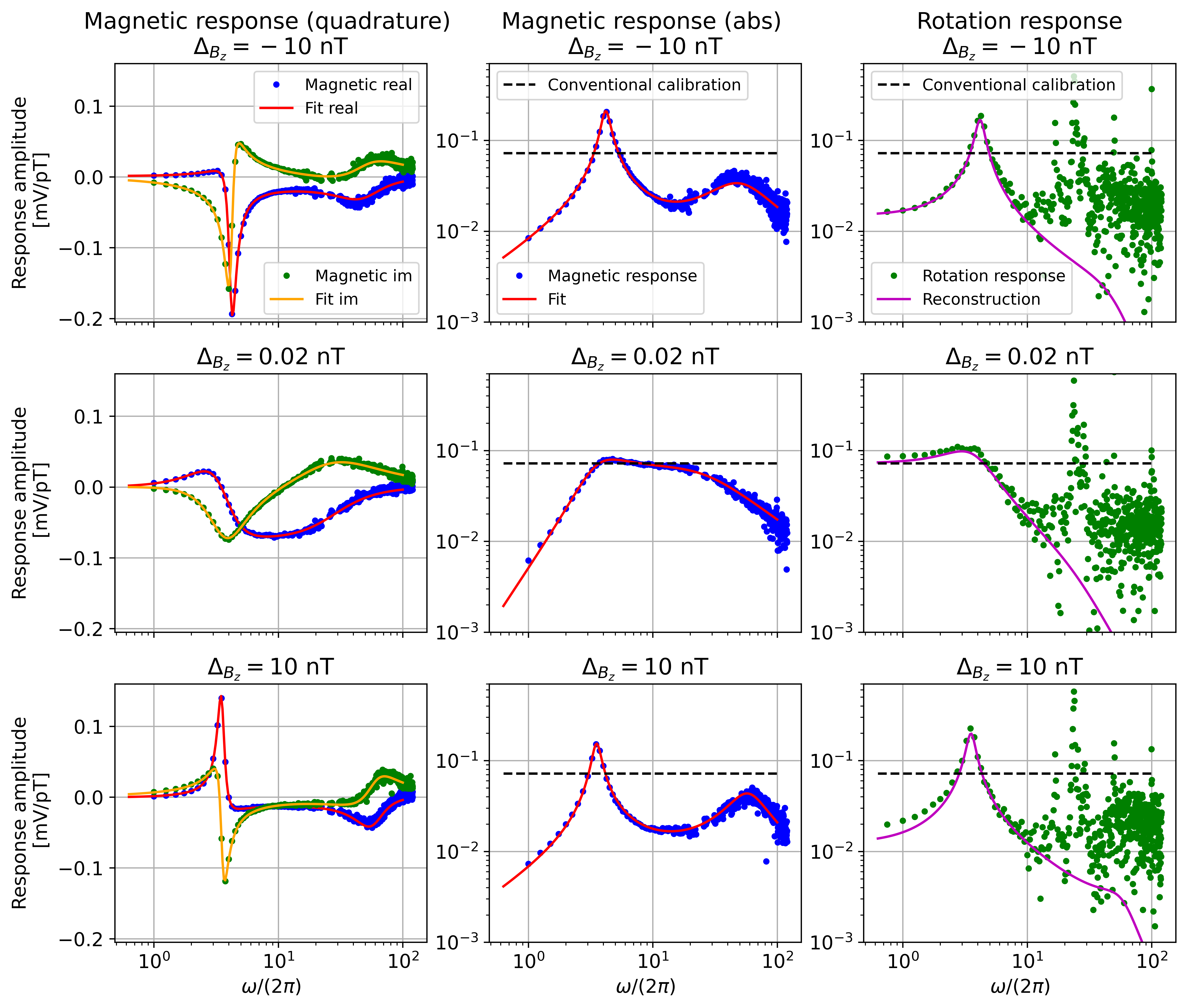}
      \caption{Frequency response to magnetic and rotation perturbation for three different values of the magnetic field detuning $\Delta_{B_z}$. The $y$ axis shows the comagnetometer response in voltage (in mV) normalized to the applied perturbation (in pT). The first and third rows are below and above the compensation point, respectively, while the middle row is very close to the compensation point. The magnetic frequency response is fitted for each magnetic detuning. The resulting parameters are used to predict the response to the rotation frequency [Eq. \eqref{eq:reconstruction_rotarion_fields}], which is then compared to the measured rotation frequency response, shown in the third column. The prediction shows excellent agreement up to about 15\,Hz, beyond which acoustic resonances of the setup and the noise floor dominate the spectrum. The dashed lines illustrate the conventional constant frequency response estimation, see for example \cite{kornack_test_nodate}.} \label{fig:fitting_and_rotation_diff_detunings}
\end{figure*}

After the step perturbation, the time series data were numerically differentiated, Fourier transformed, and divided by the (known) transfer function of the lock-in amplifier to obtain the frequency response. The unit of the frequency response is given in volts per tesla. Tesla refers here to the unit of the effective magnetic field and is used for all perturbations listed in Table\,\ref{tab:coupling_parameters}. 
The magnetic frequency response obtained for each tested value of the leading field was then fitted with the magnetic-field response model [Eq.~\eqref{fig:fitting_and_rotation_diff_detunings}].

The fitting results, along with the experimental data obtained for the time series shown in Fig.\,\ref{fig:temporal_response}, are presented in the first and second columns of Fig.\,\ref{fig:fitting_and_rotation_diff_detunings}. The fitting routine, based on complex functions, accurately captures both the phase and amplitude responses of the system, as can be seen in the plots shown in the first column of Fig.\,\ref{fig:fitting_and_rotation_diff_detunings}. The second column shows the fitted amplitude response in a log-log plot to illustrate the good agreement over the full range of frequencies. 

\begin{figure*}
    \centering
    \includegraphics[width = \textwidth]{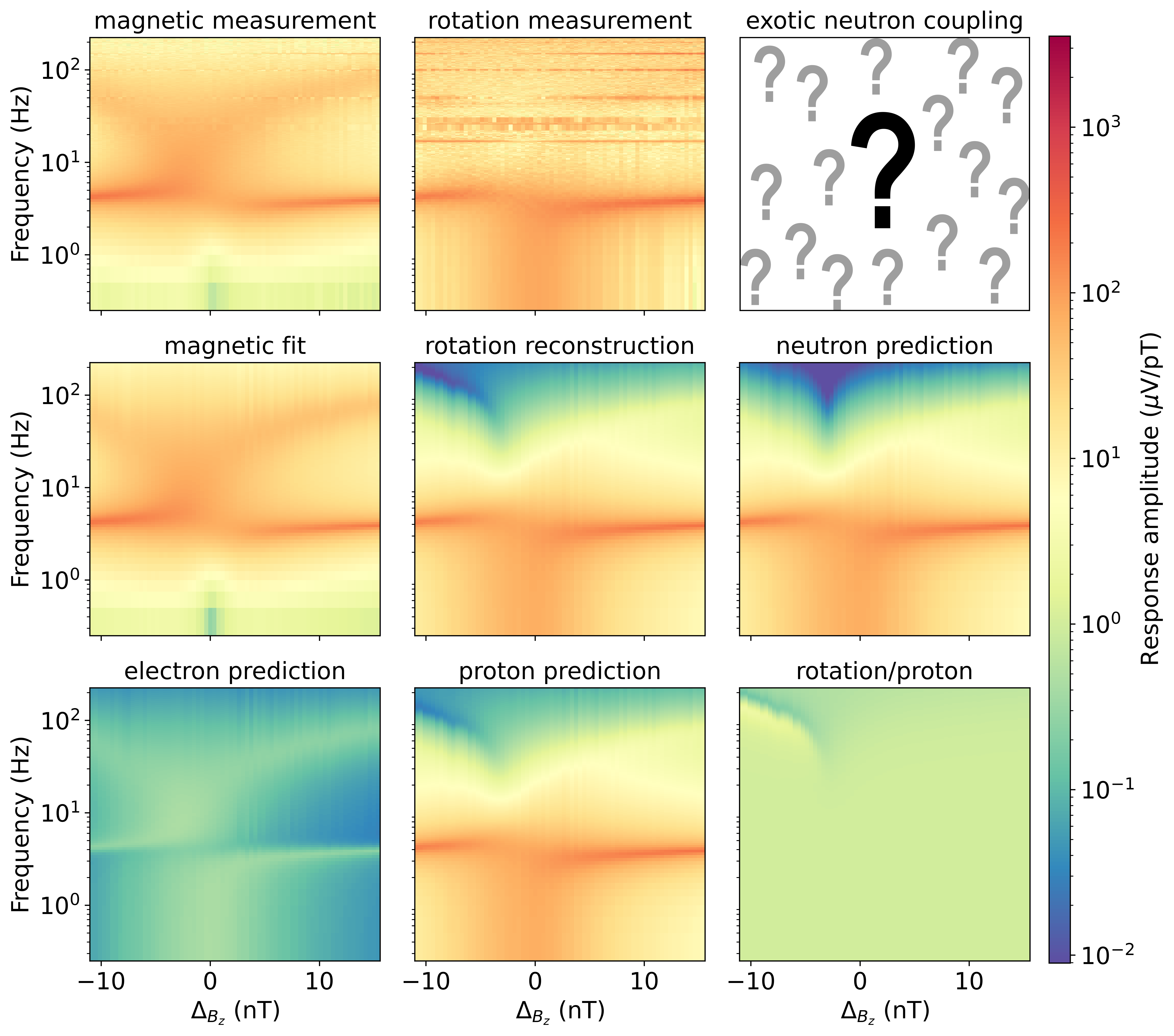}
    \caption{Directly measured frequency responses of the comagnetometer to magnetic field and rotation, along with the fit of the magnetic response and the reconstructed responses to rotation and exotic spin perturbations. The reconstructed responses were estimated on the basis of the results of the magnetic field response fitting. 
 The plots present the results obtained for different magnetic field detunings from the compensation point $\Delta_{B_z}$ with the compensation point at $\approx -120$~nT.
    The first row presented the results of the directly measured response to magnetic field and rotation. The question mark in the third column represents the absence of experimental data for yet unknown exotic spin couplings. The first column of the second row presents the results of the magnetic response fitting. The second and third column in the second row presents the results of the estimated response to rotation and neutron perturbation. The third column presents the predicted responses to electron exotic perturbation, proton exotic perturbation and the ratio between proton and rotation response.
    The color bar for all maps is expressed in $\mu$V/pT with the exception of the map for the ratio between rotation and proton coupling. Here, the numerical scale is the same as in the other plots, but the presented quantity is unitless.}
    \label{fig:map_with_results}
\end{figure*}

The third column in Fig.\,\ref{fig:fitting_and_rotation_diff_detunings} presents the directly measured frequency response to rotations and compares the data with the reconstructed response based on the fitting parameters derived from the measured magnetic response and Eqs.\,\eqref{eq:reconstruction_primary_axis} and \eqref{eq:reconstruction_rotarion_fields}. The rotation response reconstructed from the magnetic field response agrees well with the experimental data in the range from DC to 15~Hz. For higher frequencies, the experimental spectrum is masked by a series of resonances before being dominated by the noise floor of the instrument. We verified that these resonances are related to mechanical vibration modes of the magnetic shield assembly triggered by a sudden stop of rotation. 
The complete set of the results for magnetic field and rotation responses is presented as color maps in Fig.\,\ref{fig:map_with_results}. As shown, the reconstructed rotation response agrees qualitatively and quantitatively with all visible features of the rotation response measurement up to around 15\,Hz for all magnetic detunings from the compensation point. The observation that the measurement of the magnetic frequency response enables us to infer the frequency response to all other spin perturbations is the key result of this work. The protocol for this method is summarized in the flowchart in Fig.\,\ref{fig:temporal_response}. Apart from rotation sensing, this also plays an important role in determining the sensitivity of dark-matter experiments and has, to the best of our knowledge, not been fully implemented in any experiment so far. As an example of the generality of this method, we provide the inferred response of the measured comagnetometer system to exotic spin couplings of neutrons, protons, and electrons in Fig.\,\ref{fig:map_with_results}. 

The presented color maps provide insights into the dynamics of the coupled evolution of noble gas and alkali-metal polarizations. 

For large detunings from the compensation point, the Larmor resonances of both electron and nuclear polarizations are well resolved and their center frequencies depend linearly on the applied magnetic field. However, near the compensation point, the strong interaction between the polarizations merges both resonances into a hybrid response. 
In particular, one can note the canonical self-compensating mechanism: around the compensation point, the response to low-frequency magnetic perturbations is minimal while the amplitude of the response to non-magnetic couplings is maximal.
It is of interest to note that the hybrid electron-nuclear resonance occurs at a magnetic field not precisely at the compensation point but rather at $\Delta_{B_z}\approx -2$~nT, where the effective magnetic field experienced by the electron polarization crosses zero.

The predicted response for exotic proton coupling (third row, middle plot) is similar to the response to rotations [Fig.\,\ref{fig:map_with_results} (second column, row two and three)]. This similarity arises due to the similar relative coupling strength between the alkali-metal and noble-gas polarizations for both interactions (see Table\,\ref{tab:coupling_parameters}). The bottom right plot of Fig.\,\ref{fig:map_with_results} shows the ratio between the proton coupling and the rotation response. The ratio differs from unity only at high frequency and large detunings. For the chosen alkali species, the exotic coupling to neutrons affects only the noble-gas spins. This results in a difference in the response of these two couplings visible at high frequencies. The bottom left plot in Fig.~\ref{fig:map_with_results} shows the inferred frequency-dependent response of the comagnetometer to oscillating exotic couplings to electrons' polarization. Clearly visible are the Larmor resonances of the electron spin for large absolute detunings, the broad resonance width, indicating the strong damping of the electron polarization, the nuclear resonances that become visible due to being driven by the modulated electron polarization, and the avoided crossing between electron and nuclear polarization close to zero detuning. These features are similar to those observed in the magnetic response. 

\begin{figure}
    \centering\includegraphics{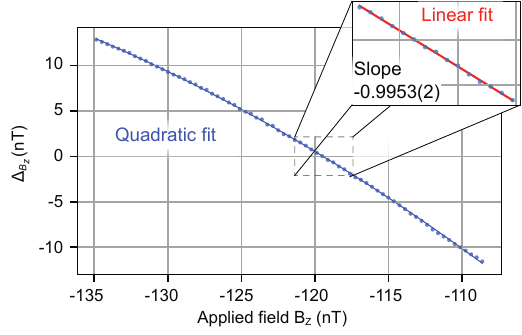}
    \caption{Fit parameter $\Delta_{ B_z}$ as a function of the applied external field $B_z$. The data are fitted with a quadratic function, which reproduces the data well. The inset shows the magnified central region around the compensation point, where the data are fitted with a linear function. The slope of this function is close to $-1$. These two parameters are completely independent and therefore an important indication for the quantitative agreement between model and experiment.}
    \label{fig:results_for_detuning}
\end{figure}

The presented calibration method provides information about the detuning from the compensation point, since it is one of the fitting parameters ($\Delta_{B_z}$) in the model, as seen in Eq.~\eqref{eq:rotation_field_fitting_model}.  Figure~\ref{fig:results_for_detuning} presents the results for the fit parameter $\Delta_{B_z}$ as a function of the applied leading field $B_z$. 
For small detunings from the compensation point ($|\Delta_{B_z}|\leq 2.5$~nT), the experimental results scale linearly with the applied leading field.
The data are fitted linearly around the compensation field, shown in the inset, resulting in a slope value of $-0.9953(2)$. This means that the leading field derived from the current through the coil and the dynamics of the comagnetometer are in agreement. This is an important cross-check on the correct calibration of the system and the fit parameters. For larger detunings ($|\Delta_{B_z}|\geq 2.5$~nT), the slope deviates from the linear behaviour and is well described by a quadratic function (see fit in Fig.~\ref{fig:results_for_detuning}). We associate this deviation with a drift of the nuclear polarization due to the leading field change that in turn alters the equilibrium nuclear polarization. This phenomenon is discussed in Ref.\,\cite{klingerPRA2023}.

\section{Conclusion\label{sec:Conclusions}}
Here we have described a calibration procedure to reliably predict the frequency-dependent response of a noble-gas-alkali-metal comagnetometer to any spin perturbation, be it magnetic or non-magnetic in nature. The calibration procedure utilizes measurements of the response of the comagnetometer system to magnetic perturbations in conjunction with a fit to a multi-parameter model based on the Bloch equations as described in Ref.\,\cite{padniuk2022response} and presented here in Eq.\,\eqref{eq:complex_fit_model}. The accuracy of the procedure was experimentally verified by successfully using it to predict the comagnetometer response to rotations.
The predicted frequency-dependent response of the comagnetometer to rotations agreed well with the directly measured comagnetometer response to rotations for a wide range of detunings from the compensation point. 
The calibration procedure is valid as long as the system response remains in the linear regime (small angles between alkali and noble-gas polarizations and leading field). The method is useful for applications of self-compensating comagnetometers both in fundamental physics tests and applied quantum gyroscopy, particularly due the technical simplicity of applying well-controlled magnetic perturbations to a comagnetometer. 

The demonstrated calibration routine outperforms calibration methods using a constant calibration factor in multiple ways. In particular, it provides the complete frequency response to any kind of spin perturbation rather than a single number that is estimated for the whole frequency-range with the other method. This is very important difference, because, as shown in Fig.\,\ref{fig:fitting_and_rotation_diff_detunings}, the frequency response is far from uniform, especially for frequencies beyond the nuclear Larmor frequency and for magnetic fields away from the compensation point (a condition that can easily occur during the course of an long-term experiment due to drifts). It should be stressed that false estimation of the frequency response to exotic couplings results in false exotic-field parameters in case of discovery and misplaced limits for a null result. For gyroscopy the consequences of an under- or overestimation of the measurement results in navigation errors with potentially catastrophic consequences. Furthermore, the demonstrated calibration method requires no changes to the main system parameters. A small step in the transverse magnetic field and a few seconds of data acquisition suffice. The system remains in equilibrium working conditions during the entire calibration procedure. This is of particular importance for long-term searches for exotic physics and navigation tasks for two reasons. First, the system requires a long time to reach stable working conditions. If the calibration protocol results in a loss of nuclear magnetization it can take hours to get back to equilibrium. Second, the calibration protocol is fast. The strong damping at the compensation point allows acquisition of all required data within seconds. This maximizes the duty cycle and the up-time of the system. And last but not least, the demonstrated calibration method determines a complete set of system parameters. This can be used to monitor the system performance over time and control the parameters in a feedback loop. This way, for example, the magnetic field detuning can be stabilized.

An immediate application of this calibration procedure is to the Advanced GNOME experiment, which will utilize a global network of comagnetometers to search for evidence of exotic physics. The calibration procedure will enable reliable long-term operation of the comagnetometer network to produce well-characterized data that can be used to search for a variety of transient signals heralding beyond-the-Standard-Model physics \cite{afach_what_2023}.

\section*{Acknowledgements}
 This work was supported by the German Federal Ministry of Education and Research (BMBF) within the Quantentechnologien program (FKZ 13N15064), by the Cluster of Excellence ``Precision Physics, Fundamental Interactions, and Structure of Matter'' (PRISMA+ EXC 2118/1) funded by the German Research Foundation (DFG) within the German Excellence Strategy (Project ID 39083149) and COST Action COSMIC WISPers CA21106, supported by COST (European Cooperation in Science and Technology). The work of D.F.J.K. was supported by U.S. National Science Foundation (NSF) Grant No. PHY-2110388. SP, MP and GL acknowledge support by the National Science Centre, Poland within the OPUS program (2020/39/B/ST2/01524).

\bibliography{Zotero_ref_corrected, ComagLitt}

\begin{thebibliography}{64}%
\makeatletter
\providecommand \@ifxundefined [1]{%
 \@ifx{#1\undefined}
}%
\providecommand \@ifnum [1]{%
 \ifnum #1\expandafter \@firstoftwo
 \else \expandafter \@secondoftwo
 \fi
}%
\providecommand \@ifx [1]{%
 \ifx #1\expandafter \@firstoftwo
 \else \expandafter \@secondoftwo
 \fi
}%
\providecommand \natexlab [1]{#1}%
\providecommand \enquote  [1]{``#1''}%
\providecommand \bibnamefont  [1]{#1}%
\providecommand \bibfnamefont [1]{#1}%
\providecommand \citenamefont [1]{#1}%
\providecommand \href@noop [0]{\@secondoftwo}%
\providecommand \href [0]{\begingroup \@sanitize@url \@href}%
\providecommand \@href[1]{\@@startlink{#1}\@@href}%
\providecommand \@@href[1]{\endgroup#1\@@endlink}%
\providecommand \@sanitize@url [0]{\catcode `\\12\catcode `\$12\catcode
  `\&12\catcode `\#12\catcode `\^12\catcode `\_12\catcode `\%12\relax}%
\providecommand \@@startlink[1]{}%
\providecommand \@@endlink[0]{}%
\providecommand \url  [0]{\begingroup\@sanitize@url \@url }%
\providecommand \@url [1]{\endgroup\@href {#1}{\urlprefix }}%
\providecommand \urlprefix  [0]{URL }%
\providecommand \Eprint [0]{\href }%
\providecommand \doibase [0]{https://doi.org/}%
\providecommand \selectlanguage [0]{\@gobble}%
\providecommand \bibinfo  [0]{\@secondoftwo}%
\providecommand \bibfield  [0]{\@secondoftwo}%
\providecommand \translation [1]{[#1]}%
\providecommand \BibitemOpen [0]{}%
\providecommand \bibitemStop [0]{}%
\providecommand \bibitemNoStop [0]{.\EOS\space}%
\providecommand \EOS [0]{\spacefactor3000\relax}%
\providecommand \BibitemShut  [1]{\csname bibitem#1\endcsname}%
\let\auto@bib@innerbib\@empty
\bibitem [{\citenamefont {Vasilakis}\ \emph {et~al.}(2009)\citenamefont
  {Vasilakis}, \citenamefont {Brown}, \citenamefont {Kornack},\ and\
  \citenamefont {Romalis}}]{vasilakis2009limits}%
  \BibitemOpen
  \bibfield  {author} {\bibinfo {author} {\bibfnamefont {G.}~\bibnamefont
  {Vasilakis}}, \bibinfo {author} {\bibfnamefont {J.~M.}\ \bibnamefont
  {Brown}}, \bibinfo {author} {\bibfnamefont {T.~W.}\ \bibnamefont {Kornack}},\
  and\ \bibinfo {author} {\bibfnamefont {M.~V.}\ \bibnamefont {Romalis}},\
  }\bibfield  {title} {\bibinfo {title} {Limits on new long range nuclear
  spin-dependent forces set with a
  $\mathbf{K}\mathrm{\text{\ensuremath{-}}}^{3}\mathrm{He}$ comagnetometer},\
  }\href {https://doi.org/10.1103/PhysRevLett.103.261801} {\bibfield  {journal}
  {\bibinfo  {journal} {Phys. Rev. Lett.}\ }\textbf {\bibinfo {volume} {103}},\
  \bibinfo {pages} {261801} (\bibinfo {year} {2009})}\BibitemShut {NoStop}%
\bibitem [{\citenamefont {Terrano}\ and\ \citenamefont
  {Romalis}(2021)}]{terrano2021comagnetometer}%
  \BibitemOpen
  \bibfield  {author} {\bibinfo {author} {\bibfnamefont {W.}~\bibnamefont
  {Terrano}}\ and\ \bibinfo {author} {\bibfnamefont {M.}~\bibnamefont
  {Romalis}},\ }\bibfield  {title} {\bibinfo {title} {Comagnetometer probes of
  dark matter and new physics},\ }\href
  {https://doi.org/10.1088/2058-9565/ac1ae0} {\bibfield  {journal} {\bibinfo
  {journal} {Quantum Sci. Technol.}\ }\textbf {\bibinfo {volume} {7}},\
  \bibinfo {pages} {014001} (\bibinfo {year} {2021})}\BibitemShut {NoStop}%
\bibitem [{\citenamefont {Bloch}\ \emph {et~al.}(2020)\citenamefont {Bloch},
  \citenamefont {Hochberg}, \citenamefont {Kuflik},\ and\ \citenamefont
  {Volansky}}]{blocholdcomag}%
  \BibitemOpen
  \bibfield  {author} {\bibinfo {author} {\bibfnamefont {I.~M.}\ \bibnamefont
  {Bloch}}, \bibinfo {author} {\bibfnamefont {Y.}~\bibnamefont {Hochberg}},
  \bibinfo {author} {\bibfnamefont {E.}~\bibnamefont {Kuflik}},\ and\ \bibinfo
  {author} {\bibfnamefont {T.}~\bibnamefont {Volansky}},\ }\bibfield  {title}
  {\bibinfo {title} {Axion-like relics: new constraints from old comagnetometer
  data},\ }\href {https://doi.org/10.1007/JHEP01(2020)167} {\bibfield
  {journal} {\bibinfo  {journal} {J. High Energy Phys.}\ }\textbf {\bibinfo
  {volume} {2020}}\bibinfo  {number} { (1)},\ \bibinfo {pages}
  {167}}\BibitemShut {NoStop}%
\bibitem [{\citenamefont {Wei}\ \emph {et~al.}(2023{\natexlab{a}})\citenamefont
  {Wei}, \citenamefont {Zhao}, \citenamefont {Fang}, \citenamefont {Xu},
  \citenamefont {Liu}, \citenamefont {Cao}, \citenamefont {Wickenbrock},
  \citenamefont {Hu}, \citenamefont {Ji}, \citenamefont {Fang},\ and\
  \citenamefont {Budker}}]{wei2022ultrasensitive}%
  \BibitemOpen
\bibfield  {number} {  }\bibfield  {author} {\bibinfo {author} {\bibfnamefont
  {K.}~\bibnamefont {Wei}}, \bibinfo {author} {\bibfnamefont {T.}~\bibnamefont
  {Zhao}}, \bibinfo {author} {\bibfnamefont {X.}~\bibnamefont {Fang}}, \bibinfo
  {author} {\bibfnamefont {Z.}~\bibnamefont {Xu}}, \bibinfo {author}
  {\bibfnamefont {C.}~\bibnamefont {Liu}}, \bibinfo {author} {\bibfnamefont
  {Q.}~\bibnamefont {Cao}}, \bibinfo {author} {\bibfnamefont {A.}~\bibnamefont
  {Wickenbrock}}, \bibinfo {author} {\bibfnamefont {Y.}~\bibnamefont {Hu}},
  \bibinfo {author} {\bibfnamefont {W.}~\bibnamefont {Ji}}, \bibinfo {author}
  {\bibfnamefont {J.}~\bibnamefont {Fang}},\ and\ \bibinfo {author}
  {\bibfnamefont {D.}~\bibnamefont {Budker}},\ }\bibfield  {title} {\bibinfo
  {title} {Ultrasensitive atomic comagnetometer with enhanced nuclear spin
  coherence},\ }\href {https://doi.org/10.1103/PhysRevLett.130.063201}
  {\bibfield  {journal} {\bibinfo  {journal} {Phys. Rev. Lett.}\ }\textbf
  {\bibinfo {volume} {130}},\ \bibinfo {pages} {063201} (\bibinfo {year}
  {2023}{\natexlab{a}})}\BibitemShut {NoStop}%
\bibitem [{\citenamefont {Kornack}\ \emph
  {et~al.}(2005{\natexlab{a}})\citenamefont {Kornack}, \citenamefont {Ghosh},\
  and\ \citenamefont {Romalis}}]{kornack2005nuclear}%
  \BibitemOpen
  \bibfield  {author} {\bibinfo {author} {\bibfnamefont {T.~W.}\ \bibnamefont
  {Kornack}}, \bibinfo {author} {\bibfnamefont {R.~K.}\ \bibnamefont {Ghosh}},\
  and\ \bibinfo {author} {\bibfnamefont {M.~V.}\ \bibnamefont {Romalis}},\
  }\bibfield  {title} {\bibinfo {title} {Nuclear spin gyroscope based on an
  atomic comagnetometer},\ }\href
  {https://doi.org/10.1103/PhysRevLett.95.230801} {\bibfield  {journal}
  {\bibinfo  {journal} {Phys. Rev. Lett.}\ }\textbf {\bibinfo {volume} {95}},\
  \bibinfo {pages} {230801} (\bibinfo {year} {2005}{\natexlab{a}})}\BibitemShut
  {NoStop}%
\bibitem [{\citenamefont {Jiang}\ \emph {et~al.}(2018)\citenamefont {Jiang},
  \citenamefont {Quan}, \citenamefont {Li}, \citenamefont {Fan}, \citenamefont
  {Liu}, \citenamefont {Qin}, \citenamefont {Wan},\ and\ \citenamefont
  {Fang}}]{jiang2018parametrically}%
  \BibitemOpen
  \bibfield  {author} {\bibinfo {author} {\bibfnamefont {L.}~\bibnamefont
  {Jiang}}, \bibinfo {author} {\bibfnamefont {W.}~\bibnamefont {Quan}},
  \bibinfo {author} {\bibfnamefont {R.}~\bibnamefont {Li}}, \bibinfo {author}
  {\bibfnamefont {W.}~\bibnamefont {Fan}}, \bibinfo {author} {\bibfnamefont
  {F.}~\bibnamefont {Liu}}, \bibinfo {author} {\bibfnamefont {J.}~\bibnamefont
  {Qin}}, \bibinfo {author} {\bibfnamefont {S.}~\bibnamefont {Wan}},\ and\
  \bibinfo {author} {\bibfnamefont {J.}~\bibnamefont {Fang}},\ }\bibfield
  {title} {\bibinfo {title} {A parametrically modulated dual-axis atomic spin
  gyroscope},\ }\href {https://doi.org/10.1063/1.5018015} {\bibfield  {journal}
  {\bibinfo  {journal} {Appl. Phys. Lett.}\ }\textbf {\bibinfo {volume}
  {112}},\ \bibinfo {pages} {054103} (\bibinfo {year} {2018})}\BibitemShut
  {NoStop}%
\bibitem [{\citenamefont {Liu}\ \emph {et~al.}(2022)\citenamefont {Liu},
  \citenamefont {Jiang}, \citenamefont {Liang}, \citenamefont {Li},
  \citenamefont {Cai}, \citenamefont {Wu},\ and\ \citenamefont
  {Quan}}]{Liu2022Comag}%
  \BibitemOpen
  \bibfield  {author} {\bibinfo {author} {\bibfnamefont {J.}~\bibnamefont
  {Liu}}, \bibinfo {author} {\bibfnamefont {L.}~\bibnamefont {Jiang}}, \bibinfo
  {author} {\bibfnamefont {Y.}~\bibnamefont {Liang}}, \bibinfo {author}
  {\bibfnamefont {G.}~\bibnamefont {Li}}, \bibinfo {author} {\bibfnamefont
  {Z.}~\bibnamefont {Cai}}, \bibinfo {author} {\bibfnamefont {Z.}~\bibnamefont
  {Wu}},\ and\ \bibinfo {author} {\bibfnamefont {W.}~\bibnamefont {Quan}},\
  }\bibfield  {title} {\bibinfo {title} {Dynamics of a spin-exchange
  relaxation-free comagnetometer for rotation sensing},\ }\href
  {https://doi.org/10.1103/PhysRevApplied.17.014030} {\bibfield  {journal}
  {\bibinfo  {journal} {Phys. Rev. Applied}\ }\textbf {\bibinfo {volume}
  {17}},\ \bibinfo {pages} {014030} (\bibinfo {year} {2022})}\BibitemShut
  {NoStop}%
\bibitem [{\citenamefont {Liang}\ \emph {et~al.}(2022)\citenamefont {Liang},
  \citenamefont {Jiang}, \citenamefont {Liu}, \citenamefont {Fan},
  \citenamefont {Zhang}, \citenamefont {Fan}, \citenamefont {Quan},\ and\
  \citenamefont {Fang}}]{Liang2022Biaxial}%
  \BibitemOpen
  \bibfield  {author} {\bibinfo {author} {\bibfnamefont {Y.}~\bibnamefont
  {Liang}}, \bibinfo {author} {\bibfnamefont {L.}~\bibnamefont {Jiang}},
  \bibinfo {author} {\bibfnamefont {J.}~\bibnamefont {Liu}}, \bibinfo {author}
  {\bibfnamefont {W.}~\bibnamefont {Fan}}, \bibinfo {author} {\bibfnamefont
  {W.}~\bibnamefont {Zhang}}, \bibinfo {author} {\bibfnamefont
  {S.}~\bibnamefont {Fan}}, \bibinfo {author} {\bibfnamefont {W.}~\bibnamefont
  {Quan}},\ and\ \bibinfo {author} {\bibfnamefont {J.}~\bibnamefont {Fang}},\
  }\bibfield  {title} {\bibinfo {title} {Biaxial signal decoupling method for
  the longitudinal magnetic-field-modulated spin-exchange-relaxation-free
  comagnetometer in inertial rotation measurement},\ }\href
  {https://doi.org/10.1103/PhysRevApplied.17.024004} {\bibfield  {journal}
  {\bibinfo  {journal} {Phys. Rev. Applied}\ }\textbf {\bibinfo {volume}
  {17}},\ \bibinfo {pages} {024004} (\bibinfo {year} {2022})}\BibitemShut
  {NoStop}%
\bibitem [{\citenamefont {Katz}\ \emph
  {et~al.}(2022{\natexlab{a}})\citenamefont {Katz}, \citenamefont {Shaham},\
  and\ \citenamefont {Firstenberg}}]{katz2022quantum}%
  \BibitemOpen
  \bibfield  {author} {\bibinfo {author} {\bibfnamefont {O.}~\bibnamefont
  {Katz}}, \bibinfo {author} {\bibfnamefont {R.}~\bibnamefont {Shaham}},\ and\
  \bibinfo {author} {\bibfnamefont {O.}~\bibnamefont {Firstenberg}},\
  }\bibfield  {title} {\bibinfo {title} {Quantum interface for noble-gas spins
  based on spin-exchange collisions},\ }\href
  {https://doi.org/10.1103/PRXQuantum.3.010305} {\bibfield  {journal} {\bibinfo
   {journal} {PRX Quantum}\ }\textbf {\bibinfo {volume} {3}},\ \bibinfo {pages}
  {010305} (\bibinfo {year} {2022}{\natexlab{a}})}\BibitemShut {NoStop}%
\bibitem [{\citenamefont {Shaham}\ \emph {et~al.}(2022)\citenamefont {Shaham},
  \citenamefont {Katz},\ and\ \citenamefont {Firstenberg}}]{shaham2022strong}%
  \BibitemOpen
  \bibfield  {author} {\bibinfo {author} {\bibfnamefont {R.}~\bibnamefont
  {Shaham}}, \bibinfo {author} {\bibfnamefont {O.}~\bibnamefont {Katz}},\ and\
  \bibinfo {author} {\bibfnamefont {O.}~\bibnamefont {Firstenberg}},\
  }\bibfield  {title} {\bibinfo {title} {Strong coupling of alkali-metal spins
  to noble-gas spins with an hour-long coherence time},\ }\href
  {https://www.nature.com/articles/s41567-022-01535-w} {\bibfield  {journal}
  {\bibinfo  {journal} {Nat. Phys.}\ }\textbf {\bibinfo {volume} {18}},\
  \bibinfo {pages} {506} (\bibinfo {year} {2022})}\BibitemShut {NoStop}%
\bibitem [{\citenamefont {Katz}\ \emph
  {et~al.}(2022{\natexlab{b}})\citenamefont {Katz}, \citenamefont {Shaham},
  \citenamefont {Reches}, \citenamefont {Gorshkov},\ and\ \citenamefont
  {Firstenberg}}]{katz2022optical}%
  \BibitemOpen
  \bibfield  {author} {\bibinfo {author} {\bibfnamefont {O.}~\bibnamefont
  {Katz}}, \bibinfo {author} {\bibfnamefont {R.}~\bibnamefont {Shaham}},
  \bibinfo {author} {\bibfnamefont {E.}~\bibnamefont {Reches}}, \bibinfo
  {author} {\bibfnamefont {A.~V.}\ \bibnamefont {Gorshkov}},\ and\ \bibinfo
  {author} {\bibfnamefont {O.}~\bibnamefont {Firstenberg}},\ }\bibfield
  {title} {\bibinfo {title} {Optical quantum memory for noble-gas spins based
  on spin-exchange collisions},\ }\href
  {https://doi.org/10.1103/PhysRevA.105.042606} {\bibfield  {journal} {\bibinfo
   {journal} {Phys. Rev. A}\ }\textbf {\bibinfo {volume} {105}},\ \bibinfo
  {pages} {042606} (\bibinfo {year} {2022}{\natexlab{b}})}\BibitemShut
  {NoStop}%
\bibitem [{\citenamefont {Walker}\ and\ \citenamefont
  {Happer}(1997)}]{walker1997spin}%
  \BibitemOpen
  \bibfield  {author} {\bibinfo {author} {\bibfnamefont {T.~G.}\ \bibnamefont
  {Walker}}\ and\ \bibinfo {author} {\bibfnamefont {W.}~\bibnamefont
  {Happer}},\ }\bibfield  {title} {\bibinfo {title} {Spin-exchange optical
  pumping of noble-gas nuclei},\ }\href
  {https://doi.org/10.1103/RevModPhys.69.629} {\bibfield  {journal} {\bibinfo
  {journal} {Rev. Mod. Phys.}\ }\textbf {\bibinfo {volume} {69}},\ \bibinfo
  {pages} {629} (\bibinfo {year} {1997})}\BibitemShut {NoStop}%
\bibitem [{\citenamefont {Jackson~Kimball}\ and\ \citenamefont {van
  Bibber}(2022)}]{kimball2022search}%
  \BibitemOpen
  \bibfield  {author} {\bibinfo {author} {\bibfnamefont {D.~F.}\ \bibnamefont
  {Jackson~Kimball}}\ and\ \bibinfo {author} {\bibfnamefont {K.}~\bibnamefont
  {van Bibber}},\ }\href {https://doi.org/10.1007/978-3-030-95852-7} {\emph
  {\bibinfo {title} {{The Search for Ultralight Bosonic Dark Matter}}}}\
  (\bibinfo  {publisher} {Springer},\ \bibinfo {year} {2022})\BibitemShut
  {NoStop}%
\bibitem [{\citenamefont {Chupp}\ \emph {et~al.}(2019)\citenamefont {Chupp},
  \citenamefont {Fierlinger}, \citenamefont {Ramsey-Musolf},\ and\
  \citenamefont {Singh}}]{chupp2019electric}%
  \BibitemOpen
  \bibfield  {author} {\bibinfo {author} {\bibfnamefont {T.}~\bibnamefont
  {Chupp}}, \bibinfo {author} {\bibfnamefont {P.}~\bibnamefont {Fierlinger}},
  \bibinfo {author} {\bibfnamefont {M.}~\bibnamefont {Ramsey-Musolf}},\ and\
  \bibinfo {author} {\bibfnamefont {J.}~\bibnamefont {Singh}},\ }\bibfield
  {title} {\bibinfo {title} {Electric dipole moments of atoms, molecules,
  nuclei, and particles},\ }\href
  {https://doi.org/10.1103/RevModPhys.91.015001} {\bibfield  {journal}
  {\bibinfo  {journal} {Rev. Mod. Phys.}\ }\textbf {\bibinfo {volume} {91}},\
  \bibinfo {pages} {015001} (\bibinfo {year} {2019})}\BibitemShut {NoStop}%
\bibitem [{\citenamefont {Fang}\ and\ \citenamefont
  {Qin}(2012)}]{fang2012advances}%
  \BibitemOpen
  \bibfield  {author} {\bibinfo {author} {\bibfnamefont {J.}~\bibnamefont
  {Fang}}\ and\ \bibinfo {author} {\bibfnamefont {J.}~\bibnamefont {Qin}},\
  }\bibfield  {title} {\bibinfo {title} {Advances in atomic gyroscopes: A view
  from inertial navigation applications},\ }\href
  {https://doi.org/10.3390/s120506331} {\bibfield  {journal} {\bibinfo
  {journal} {Sensors}\ }\textbf {\bibinfo {volume} {12}},\ \bibinfo {pages}
  {6331} (\bibinfo {year} {2012})}\BibitemShut {NoStop}%
\bibitem [{\citenamefont {Cai}\ \emph {et~al.}(2018)\citenamefont {Cai},
  \citenamefont {Yang}, \citenamefont {Quan}, \citenamefont {Song},
  \citenamefont {Tu},\ and\ \citenamefont {Liu}}]{cai2018error}%
  \BibitemOpen
  \bibfield  {author} {\bibinfo {author} {\bibfnamefont {Q.}~\bibnamefont
  {Cai}}, \bibinfo {author} {\bibfnamefont {G.}~\bibnamefont {Yang}}, \bibinfo
  {author} {\bibfnamefont {W.}~\bibnamefont {Quan}}, \bibinfo {author}
  {\bibfnamefont {N.}~\bibnamefont {Song}}, \bibinfo {author} {\bibfnamefont
  {Y.}~\bibnamefont {Tu}},\ and\ \bibinfo {author} {\bibfnamefont
  {Y.}~\bibnamefont {Liu}},\ }\bibfield  {title} {\bibinfo {title} {{Error
  analysis of the K-Rb-$^{21}$Ne comagnetometer space-stable inertial
  navigation system}},\ }\href {https://doi.org/10.3390/s18020670} {\bibfield
  {journal} {\bibinfo  {journal} {Sensors}\ }\textbf {\bibinfo {volume} {18}},\
  \bibinfo {pages} {670} (\bibinfo {year} {2018})}\BibitemShut {NoStop}%
\bibitem [{\citenamefont {Smiciklas}\ \emph {et~al.}(2011)\citenamefont
  {Smiciklas}, \citenamefont {Brown}, \citenamefont {Cheuk}, \citenamefont
  {Smullin},\ and\ \citenamefont {Romalis}}]{Smiciklas2011}%
  \BibitemOpen
  \bibfield  {author} {\bibinfo {author} {\bibfnamefont {M.}~\bibnamefont
  {Smiciklas}}, \bibinfo {author} {\bibfnamefont {J.~M.}\ \bibnamefont
  {Brown}}, \bibinfo {author} {\bibfnamefont {L.~W.}\ \bibnamefont {Cheuk}},
  \bibinfo {author} {\bibfnamefont {S.~J.}\ \bibnamefont {Smullin}},\ and\
  \bibinfo {author} {\bibfnamefont {M.~V.}\ \bibnamefont {Romalis}},\
  }\bibfield  {title} {\bibinfo {title} {New test of local lorentz invariance
  using a
  $^{21}\mathrm{Ne}\mathrm{\text{\ensuremath{-}}}\mathrm{Rb}\mathrm{\text{\ensuremath{-}}}\mathbf{K}$
  comagnetometer},\ }\href {https://doi.org/10.1103/PhysRevLett.107.171604}
  {\bibfield  {journal} {\bibinfo  {journal} {Phys. Rev. Lett.}\ }\textbf
  {\bibinfo {volume} {107}},\ \bibinfo {pages} {171604} (\bibinfo {year}
  {2011})}\BibitemShut {NoStop}%
\bibitem [{\citenamefont {Zhang}\ \emph {et~al.}(2023)\citenamefont {Zhang},
  \citenamefont {Ba}, \citenamefont {Ning}, \citenamefont {Zhai}, \citenamefont
  {Lu},\ and\ \citenamefont {Sheng}}]{sheng2023spingravity}%
  \BibitemOpen
  \bibfield  {author} {\bibinfo {author} {\bibfnamefont {S.-B.}\ \bibnamefont
  {Zhang}}, \bibinfo {author} {\bibfnamefont {Z.-L.}\ \bibnamefont {Ba}},
  \bibinfo {author} {\bibfnamefont {D.-H.}\ \bibnamefont {Ning}}, \bibinfo
  {author} {\bibfnamefont {N.-F.}\ \bibnamefont {Zhai}}, \bibinfo {author}
  {\bibfnamefont {Z.-T.}\ \bibnamefont {Lu}},\ and\ \bibinfo {author}
  {\bibfnamefont {D.}~\bibnamefont {Sheng}},\ }\bibfield  {title} {\bibinfo
  {title} {Search for spin-dependent gravitational interactions at earth
  range},\ }\href {https://doi.org/10.1103/PhysRevLett.130.201401} {\bibfield
  {journal} {\bibinfo  {journal} {Phys. Rev. Lett.}\ }\textbf {\bibinfo
  {volume} {130}},\ \bibinfo {pages} {201401} (\bibinfo {year}
  {2023})}\BibitemShut {NoStop}%
\bibitem [{\citenamefont {Lee}\ \emph {et~al.}(2018)\citenamefont {Lee},
  \citenamefont {Almasi},\ and\ \citenamefont {Romalis}}]{Lee2018}%
  \BibitemOpen
  \bibfield  {author} {\bibinfo {author} {\bibfnamefont {J.}~\bibnamefont
  {Lee}}, \bibinfo {author} {\bibfnamefont {A.}~\bibnamefont {Almasi}},\ and\
  \bibinfo {author} {\bibfnamefont {M.}~\bibnamefont {Romalis}},\ }\bibfield
  {title} {\bibinfo {title} {Improved limits on spin-mass interactions},\
  }\href {https://doi.org/10.1103/PhysRevLett.120.161801} {\bibfield  {journal}
  {\bibinfo  {journal} {Phys. Rev. Lett.}\ }\textbf {\bibinfo {volume} {120}},\
  \bibinfo {pages} {161801} (\bibinfo {year} {2018})}\BibitemShut {NoStop}%
\bibitem [{\citenamefont {Graham}\ \emph
  {et~al.}(2015{\natexlab{a}})\citenamefont {Graham}, \citenamefont {Kaplan},\
  and\ \citenamefont {Rajendran}}]{Gra15}%
  \BibitemOpen
  \bibfield  {author} {\bibinfo {author} {\bibfnamefont {P.~W.}\ \bibnamefont
  {Graham}}, \bibinfo {author} {\bibfnamefont {D.~E.}\ \bibnamefont {Kaplan}},\
  and\ \bibinfo {author} {\bibfnamefont {S.}~\bibnamefont {Rajendran}},\
  }\bibfield  {title} {\bibinfo {title} {{Cosmological Relaxation of the
  Electroweak Scale}},\ }\href {https://doi.org/10.1103/PhysRevLett.115.221801}
  {\bibfield  {journal} {\bibinfo  {journal} {Phys. Rev. Lett.}\ }\textbf
  {\bibinfo {volume} {115}},\ \bibinfo {pages} {221801} (\bibinfo {year}
  {2015}{\natexlab{a}})}\BibitemShut {NoStop}%
\bibitem [{\citenamefont {Co}\ \emph {et~al.}(2021)\citenamefont {Co},
  \citenamefont {Hall},\ and\ \citenamefont {Harigaya}}]{co2020predictions}%
  \BibitemOpen
  \bibfield  {author} {\bibinfo {author} {\bibfnamefont {R.~T.}\ \bibnamefont
  {Co}}, \bibinfo {author} {\bibfnamefont {L.~J.}\ \bibnamefont {Hall}},\ and\
  \bibinfo {author} {\bibfnamefont {K.}~\bibnamefont {Harigaya}},\ }\bibfield
  {title} {\bibinfo {title} {{Predictions for Axion Couplings from ALP
  Cogenesis}},\ }\href {https://doi.org/10.1007/JHEP01(2021)172} {\bibfield
  {journal} {\bibinfo  {journal} {J. High Energ. Phys.}\ }\textbf {\bibinfo
  {volume} {2021}},\ \bibinfo {pages} {172}}\BibitemShut {NoStop}%
\bibitem [{\citenamefont {Svrcek}\ and\ \citenamefont {Witten}(2006)}]{Svr06}%
  \BibitemOpen
  \bibfield  {author} {\bibinfo {author} {\bibfnamefont {P.}~\bibnamefont
  {Svrcek}}\ and\ \bibinfo {author} {\bibfnamefont {E.}~\bibnamefont
  {Witten}},\ }\bibfield  {title} {\bibinfo {title} {Axions in string theory},\
  }\href {https://doi.org/10.1088/1126-6708/2006/06/051} {\bibfield  {journal}
  {\bibinfo  {journal} {J. High Energy Phys.}\ }\textbf {\bibinfo {volume}
  {2006}}\bibinfo  {number} { (06)},\ \bibinfo {pages} {051}}\BibitemShut
  {NoStop}%
\bibitem [{\citenamefont {Arvanitaki}\ \emph {et~al.}(2010)\citenamefont
  {Arvanitaki}, \citenamefont {Dimopoulos}, \citenamefont {Dubovsky},
  \citenamefont {Kaloper},\ and\ \citenamefont {March-Russell}}]{Arv10}%
  \BibitemOpen
\bibfield  {number} {  }\bibfield  {author} {\bibinfo {author} {\bibfnamefont
  {A.}~\bibnamefont {Arvanitaki}}, \bibinfo {author} {\bibfnamefont
  {S.}~\bibnamefont {Dimopoulos}}, \bibinfo {author} {\bibfnamefont
  {S.}~\bibnamefont {Dubovsky}}, \bibinfo {author} {\bibfnamefont
  {N.}~\bibnamefont {Kaloper}},\ and\ \bibinfo {author} {\bibfnamefont
  {J.}~\bibnamefont {March-Russell}},\ }\bibfield  {title} {\bibinfo {title}
  {String axiverse},\ }\href {https://doi.org/10.1103/PhysRevD.81.123530}
  {\bibfield  {journal} {\bibinfo  {journal} {Phys. Rev. D}\ }\textbf {\bibinfo
  {volume} {81}},\ \bibinfo {pages} {123530} (\bibinfo {year}
  {2010})}\BibitemShut {NoStop}%
\bibitem [{\citenamefont {Graham}\ \emph {et~al.}(2016)\citenamefont {Graham},
  \citenamefont {Kaplan}, \citenamefont {Mardon}, \citenamefont {Rajendran},\
  and\ \citenamefont {Terrano}}]{graham2016dark}%
  \BibitemOpen
  \bibfield  {author} {\bibinfo {author} {\bibfnamefont {P.~W.}\ \bibnamefont
  {Graham}}, \bibinfo {author} {\bibfnamefont {D.~E.}\ \bibnamefont {Kaplan}},
  \bibinfo {author} {\bibfnamefont {J.}~\bibnamefont {Mardon}}, \bibinfo
  {author} {\bibfnamefont {S.}~\bibnamefont {Rajendran}},\ and\ \bibinfo
  {author} {\bibfnamefont {W.~A.}\ \bibnamefont {Terrano}},\ }\bibfield
  {title} {\bibinfo {title} {Dark matter direct detection with
  accelerometers},\ }\href {https://doi.org/10.1103/PhysRevD.93.075029}
  {\bibfield  {journal} {\bibinfo  {journal} {Phys. Rev. D}\ }\textbf {\bibinfo
  {volume} {93}},\ \bibinfo {pages} {075029} (\bibinfo {year}
  {2016})}\BibitemShut {NoStop}%
\bibitem [{\citenamefont {Safronova}\ \emph {et~al.}(2018)\citenamefont
  {Safronova}, \citenamefont {Budker}, \citenamefont {DeMille}, \citenamefont
  {Jackson~Kimball}, \citenamefont {Derevianko},\ and\ \citenamefont
  {Clark}}]{safronova2018search}%
  \BibitemOpen
  \bibfield  {author} {\bibinfo {author} {\bibfnamefont {M.}~\bibnamefont
  {Safronova}}, \bibinfo {author} {\bibfnamefont {D.}~\bibnamefont {Budker}},
  \bibinfo {author} {\bibfnamefont {D.}~\bibnamefont {DeMille}}, \bibinfo
  {author} {\bibfnamefont {D.~F.}\ \bibnamefont {Jackson~Kimball}}, \bibinfo
  {author} {\bibfnamefont {A.}~\bibnamefont {Derevianko}},\ and\ \bibinfo
  {author} {\bibfnamefont {C.~W.}\ \bibnamefont {Clark}},\ }\bibfield  {title}
  {\bibinfo {title} {Search for new physics with atoms and molecules},\ }\href
  {https://doi.org/10.1103/RevModPhys.90.025008} {\bibfield  {journal}
  {\bibinfo  {journal} {Rev. Mod. Phys.}\ }\textbf {\bibinfo {volume} {90}},\
  \bibinfo {pages} {025008} (\bibinfo {year} {2018})}\BibitemShut {NoStop}%
\bibitem [{\citenamefont {Graham}\ \emph
  {et~al.}(2015{\natexlab{b}})\citenamefont {Graham}, \citenamefont
  {Irastorza}, \citenamefont {Lamoreaux}, \citenamefont {Lindner},\ and\
  \citenamefont {van Bibber}}]{graham2015experimental}%
  \BibitemOpen
  \bibfield  {author} {\bibinfo {author} {\bibfnamefont {P.~W.}\ \bibnamefont
  {Graham}}, \bibinfo {author} {\bibfnamefont {I.~G.}\ \bibnamefont
  {Irastorza}}, \bibinfo {author} {\bibfnamefont {S.~K.}\ \bibnamefont
  {Lamoreaux}}, \bibinfo {author} {\bibfnamefont {A.}~\bibnamefont {Lindner}},\
  and\ \bibinfo {author} {\bibfnamefont {K.~A.}\ \bibnamefont {van Bibber}},\
  }\bibfield  {title} {\bibinfo {title} {Experimental searches for the axion
  and axion-like particles},\ }\href
  {https://doi.org/10.1146/annurev-nucl-102014-022120} {\bibfield  {journal}
  {\bibinfo  {journal} {Annu. Rev. Nucl. Part. Sci.}\ }\textbf {\bibinfo
  {volume} {65}},\ \bibinfo {pages} {485} (\bibinfo {year}
  {2015}{\natexlab{b}})}\BibitemShut {NoStop}%
\bibitem [{\citenamefont {Graham}\ \emph {et~al.}(2018)\citenamefont {Graham},
  \citenamefont {Kaplan}, \citenamefont {Mardon}, \citenamefont {Rajendran},
  \citenamefont {Terrano}, \citenamefont {Trahms},\ and\ \citenamefont
  {Wilkason}}]{graham2018spin}%
  \BibitemOpen
  \bibfield  {author} {\bibinfo {author} {\bibfnamefont {P.~W.}\ \bibnamefont
  {Graham}}, \bibinfo {author} {\bibfnamefont {D.~E.}\ \bibnamefont {Kaplan}},
  \bibinfo {author} {\bibfnamefont {J.}~\bibnamefont {Mardon}}, \bibinfo
  {author} {\bibfnamefont {S.}~\bibnamefont {Rajendran}}, \bibinfo {author}
  {\bibfnamefont {W.~A.}\ \bibnamefont {Terrano}}, \bibinfo {author}
  {\bibfnamefont {L.}~\bibnamefont {Trahms}},\ and\ \bibinfo {author}
  {\bibfnamefont {T.}~\bibnamefont {Wilkason}},\ }\bibfield  {title} {\bibinfo
  {title} {Spin precession experiments for light axionic dark matter},\ }\href
  {https://doi.org/10.1103/PhysRevD.97.055006} {\bibfield  {journal} {\bibinfo
  {journal} {Phys. Rev. D}\ }\textbf {\bibinfo {volume} {97}},\ \bibinfo
  {pages} {055006} (\bibinfo {year} {2018})}\BibitemShut {NoStop}%
\bibitem [{\citenamefont {Afach}\ \emph {et~al.}(2023)\citenamefont {Afach},
  \citenamefont {Aybas~Tumturk}, \citenamefont {Bekker}, \citenamefont
  {Buchler}, \citenamefont {Budker}, \citenamefont {Cervantes}, \citenamefont
  {Derevianko}, \citenamefont {Eby}, \citenamefont {Figueroa}, \citenamefont
  {Folman} \emph {et~al.}}]{afach_what_2023}%
  \BibitemOpen
  \bibfield  {author} {\bibinfo {author} {\bibfnamefont {S.}~\bibnamefont
  {Afach}}, \bibinfo {author} {\bibfnamefont {D.}~\bibnamefont
  {Aybas~Tumturk}}, \bibinfo {author} {\bibfnamefont {H.}~\bibnamefont
  {Bekker}}, \bibinfo {author} {\bibfnamefont {B.~C.}\ \bibnamefont {Buchler}},
  \bibinfo {author} {\bibfnamefont {D.}~\bibnamefont {Budker}}, \bibinfo
  {author} {\bibfnamefont {K.}~\bibnamefont {Cervantes}}, \bibinfo {author}
  {\bibfnamefont {A.}~\bibnamefont {Derevianko}}, \bibinfo {author}
  {\bibfnamefont {J.}~\bibnamefont {Eby}}, \bibinfo {author} {\bibfnamefont
  {N.~L.}\ \bibnamefont {Figueroa}}, \bibinfo {author} {\bibfnamefont
  {R.}~\bibnamefont {Folman}}, \emph {et~al.},\ }\bibfield  {title} {\bibinfo
  {title} {What can a {GNOME} do? {Search} targets for the {Global} {Network}
  of {Optical} {Magnetometers} for {Exotic} physics searches},\ }\href
  {https://doi.org/10.1002/andp.202300083} {\bibfield  {journal} {\bibinfo
  {journal} {Ann. Phys. (Berlin)}\ }\textbf {\bibinfo {volume} {2023}},\
  \bibinfo {pages} {2300083} (\bibinfo {year} {2023})}\BibitemShut {NoStop}%
\bibitem [{\citenamefont {Lee}\ \emph {et~al.}(2023)\citenamefont {Lee},
  \citenamefont {Lisanti}, \citenamefont {Terrano},\ and\ \citenamefont
  {Romalis}}]{Lee2023}%
  \BibitemOpen
  \bibfield  {author} {\bibinfo {author} {\bibfnamefont {J.}~\bibnamefont
  {Lee}}, \bibinfo {author} {\bibfnamefont {M.}~\bibnamefont {Lisanti}},
  \bibinfo {author} {\bibfnamefont {W.~A.}\ \bibnamefont {Terrano}},\ and\
  \bibinfo {author} {\bibfnamefont {M.}~\bibnamefont {Romalis}},\ }\bibfield
  {title} {\bibinfo {title} {Laboratory constraints on the neutron-spin
  coupling of fev-scale axions},\ }\href
  {https://doi.org/10.1103/PhysRevX.13.011050} {\bibfield  {journal} {\bibinfo
  {journal} {Phys. Rev. X}\ }\textbf {\bibinfo {volume} {13}},\ \bibinfo
  {pages} {011050} (\bibinfo {year} {2023})}\BibitemShut {NoStop}%
\bibitem [{\citenamefont {Wei}\ \emph {et~al.}(2023{\natexlab{b}})\citenamefont
  {Wei}, \citenamefont {Xu}, \citenamefont {He}, \citenamefont {Ma},
  \citenamefont {Heng}, \citenamefont {Huang}, \citenamefont {Quan},
  \citenamefont {Ji}, \citenamefont {Liu}, \citenamefont {Wang}, \citenamefont
  {Fang},\ and\ \citenamefont {Budker}}]{wei_dark_2023}%
  \BibitemOpen
  \bibfield  {author} {\bibinfo {author} {\bibfnamefont {K.}~\bibnamefont
  {Wei}}, \bibinfo {author} {\bibfnamefont {Z.}~\bibnamefont {Xu}}, \bibinfo
  {author} {\bibfnamefont {Y.}~\bibnamefont {He}}, \bibinfo {author}
  {\bibfnamefont {X.}~\bibnamefont {Ma}}, \bibinfo {author} {\bibfnamefont
  {X.}~\bibnamefont {Heng}}, \bibinfo {author} {\bibfnamefont {X.}~\bibnamefont
  {Huang}}, \bibinfo {author} {\bibfnamefont {W.}~\bibnamefont {Quan}},
  \bibinfo {author} {\bibfnamefont {W.}~\bibnamefont {Ji}}, \bibinfo {author}
  {\bibfnamefont {J.}~\bibnamefont {Liu}}, \bibinfo {author} {\bibfnamefont
  {X.}~\bibnamefont {Wang}}, \bibinfo {author} {\bibfnamefont {J.}~\bibnamefont
  {Fang}},\ and\ \bibinfo {author} {\bibfnamefont {D.}~\bibnamefont {Budker}},\
  }\bibfield  {title} {\bibinfo {title} {Dark matter search with a
  strongly-coupled hybrid spin system},\ }\href
  {http://arxiv.org/abs/2306.08039} {\bibfield  {journal} {\bibinfo  {journal}
  {arXiv:2306.08039}\ } (\bibinfo {year} {2023}{\natexlab{b}})}\BibitemShut
  {NoStop}%
\bibitem [{\citenamefont {Bloch}\ \emph {et~al.}(2022)\citenamefont {Bloch},
  \citenamefont {Ronen}, \citenamefont {Shaham}, \citenamefont {Katz},
  \citenamefont {Volansky},\ and\ \citenamefont {Katz}}]{bloch2022New}%
  \BibitemOpen
  \bibfield  {author} {\bibinfo {author} {\bibfnamefont {I.~M.}\ \bibnamefont
  {Bloch}}, \bibinfo {author} {\bibfnamefont {G.}~\bibnamefont {Ronen}},
  \bibinfo {author} {\bibfnamefont {R.}~\bibnamefont {Shaham}}, \bibinfo
  {author} {\bibfnamefont {O.}~\bibnamefont {Katz}}, \bibinfo {author}
  {\bibfnamefont {T.}~\bibnamefont {Volansky}},\ and\ \bibinfo {author}
  {\bibfnamefont {O.}~\bibnamefont {Katz}},\ }\bibfield  {title} {\bibinfo
  {title} {New constraints on axion-like dark matter using a floquet quantum
  detector},\ }\href {https://www.science.org/doi/abs/10.1126/sciadv.abl8919}
  {\bibfield  {journal} {\bibinfo  {journal} {Sci. Adv.}\ }\textbf {\bibinfo
  {volume} {8}} (\bibinfo {year} {2022})}\BibitemShut {NoStop}%
\bibitem [{\citenamefont {Bloch}\ \emph {et~al.}(2023)\citenamefont {Bloch},
  \citenamefont {Shaham}, \citenamefont {Hochberg}, \citenamefont {Kuflik},
  \citenamefont {Volansky},\ and\ \citenamefont {Katz}}]{Bloch:2023:natComm}%
  \BibitemOpen
  \bibfield  {author} {\bibinfo {author} {\bibfnamefont {I.~M.}\ \bibnamefont
  {Bloch}}, \bibinfo {author} {\bibfnamefont {R.}~\bibnamefont {Shaham}},
  \bibinfo {author} {\bibfnamefont {Y.}~\bibnamefont {Hochberg}}, \bibinfo
  {author} {\bibfnamefont {E.}~\bibnamefont {Kuflik}}, \bibinfo {author}
  {\bibfnamefont {T.}~\bibnamefont {Volansky}},\ and\ \bibinfo {author}
  {\bibfnamefont {O.}~\bibnamefont {Katz}},\ }\bibfield  {title} {\bibinfo
  {title} {Constraints on axion-like dark matter from a {SERF}
  comagnetometer},\ }\href {https://doi.org/10.1038/s41467-023-41162-4}
  {\bibfield  {journal} {\bibinfo  {journal} {Nature Comm.}\ }\textbf {\bibinfo
  {volume} {14}},\ \bibinfo {pages} {5784} (\bibinfo {year}
  {2023})}\BibitemShut {NoStop}%
\bibitem [{\citenamefont {Centers}\ \emph {et~al.}(2021)\citenamefont
  {Centers}, \citenamefont {Blanchard}, \citenamefont {Conrad}, \citenamefont
  {Figueroa}, \citenamefont {Garcon}, \citenamefont {Gramolin}, \citenamefont
  {Jackson~Kimball}, \citenamefont {Lawson}, \citenamefont {Pelssers},
  \citenamefont {Smiga} \emph {et~al.}}]{centers2021stochastic}%
  \BibitemOpen
  \bibfield  {author} {\bibinfo {author} {\bibfnamefont {G.~P.}\ \bibnamefont
  {Centers}}, \bibinfo {author} {\bibfnamefont {J.~W.}\ \bibnamefont
  {Blanchard}}, \bibinfo {author} {\bibfnamefont {J.}~\bibnamefont {Conrad}},
  \bibinfo {author} {\bibfnamefont {N.~L.}\ \bibnamefont {Figueroa}}, \bibinfo
  {author} {\bibfnamefont {A.}~\bibnamefont {Garcon}}, \bibinfo {author}
  {\bibfnamefont {A.~V.}\ \bibnamefont {Gramolin}}, \bibinfo {author}
  {\bibfnamefont {D.~F.}\ \bibnamefont {Jackson~Kimball}}, \bibinfo {author}
  {\bibfnamefont {M.}~\bibnamefont {Lawson}}, \bibinfo {author} {\bibfnamefont
  {B.}~\bibnamefont {Pelssers}}, \bibinfo {author} {\bibfnamefont {J.~A.}\
  \bibnamefont {Smiga}}, \emph {et~al.},\ }\bibfield  {title} {\bibinfo {title}
  {Stochastic fluctuations of bosonic dark matter},\ }\href
  {https://doi.org/10.1038/s41467-021-27632-7} {\bibfield  {journal} {\bibinfo
  {journal} {Nature Comm.}\ }\textbf {\bibinfo {volume} {12}},\ \bibinfo
  {pages} {7321} (\bibinfo {year} {2021})}\BibitemShut {NoStop}%
\bibitem [{\citenamefont {Pustelny}\ \emph {et~al.}(2013)\citenamefont
  {Pustelny}, \citenamefont {Jackson~Kimball}, \citenamefont {Pankow},
  \citenamefont {Ledbetter}, \citenamefont {Wlodarczyk}, \citenamefont
  {Wcislo}, \citenamefont {Pospelov}, \citenamefont {Smith}, \citenamefont
  {Read}, \citenamefont {Gawlik},\ and\ \citenamefont
  {Budker}}]{pustelny_global_2013}%
  \BibitemOpen
  \bibfield  {author} {\bibinfo {author} {\bibfnamefont {S.}~\bibnamefont
  {Pustelny}}, \bibinfo {author} {\bibfnamefont {D.~F.}\ \bibnamefont
  {Jackson~Kimball}}, \bibinfo {author} {\bibfnamefont {C.}~\bibnamefont
  {Pankow}}, \bibinfo {author} {\bibfnamefont {M.~P.}\ \bibnamefont
  {Ledbetter}}, \bibinfo {author} {\bibfnamefont {P.}~\bibnamefont
  {Wlodarczyk}}, \bibinfo {author} {\bibfnamefont {P.}~\bibnamefont {Wcislo}},
  \bibinfo {author} {\bibfnamefont {M.}~\bibnamefont {Pospelov}}, \bibinfo
  {author} {\bibfnamefont {J.~R.}\ \bibnamefont {Smith}}, \bibinfo {author}
  {\bibfnamefont {J.}~\bibnamefont {Read}}, \bibinfo {author} {\bibfnamefont
  {W.}~\bibnamefont {Gawlik}},\ and\ \bibinfo {author} {\bibfnamefont
  {D.}~\bibnamefont {Budker}},\ }\bibfield  {title} {\bibinfo {title} {The
  {Global} {Network} of {Optical} {Magnetometers} for {Exotic} physics
  ({GNOME}): {A} novel scheme to search for physics beyond the {Standard}
  {Model}},\ }\href {https://doi.org/10.1002/andp.201300061} {\bibfield
  {journal} {\bibinfo  {journal} {Ann. Phys. (Berlin)}\ }\textbf {\bibinfo
  {volume} {525}},\ \bibinfo {pages} {659} (\bibinfo {year}
  {2013})}\BibitemShut {NoStop}%
\bibitem [{\citenamefont {Afach}\ \emph {et~al.}(2018)\citenamefont {Afach},
  \citenamefont {Budker}, \citenamefont {DeCamp}, \citenamefont {Dumont},
  \citenamefont {Grujić}, \citenamefont {Guo}, \citenamefont
  {Jackson Kimball}, \citenamefont {Kornack}, \citenamefont {Lebedev},
  \citenamefont {Li}, \citenamefont {Masia-Roig}, \citenamefont {Nix},
  \citenamefont {Padniuk}, \citenamefont {Palm}, \citenamefont {Pankow},
  \citenamefont {Penaflor}, \citenamefont {Peng}, \citenamefont {Pustelny},
  \citenamefont {Scholtes}, \citenamefont {Smiga}, \citenamefont {Stalnaker},
  \citenamefont {Weis}, \citenamefont {Wickenbrock},\ and\ \citenamefont
  {Wurm}}]{afach_characterization_2018}%
  \BibitemOpen
  \bibfield  {author} {\bibinfo {author} {\bibfnamefont {S.}~\bibnamefont
  {Afach}}, \bibinfo {author} {\bibfnamefont {D.}~\bibnamefont {Budker}},
  \bibinfo {author} {\bibfnamefont {G.}~\bibnamefont {DeCamp}}, \bibinfo
  {author} {\bibfnamefont {V.}~\bibnamefont {Dumont}}, \bibinfo {author}
  {\bibfnamefont {Z.}~\bibnamefont {Grujić}}, \bibinfo {author} {\bibfnamefont
  {H.}~\bibnamefont {Guo}}, \bibinfo {author} {\bibfnamefont {D.}~\bibnamefont
  {Jackson Kimball}}, \bibinfo {author} {\bibfnamefont {T.}~\bibnamefont
  {Kornack}}, \bibinfo {author} {\bibfnamefont {V.}~\bibnamefont {Lebedev}},
  \bibinfo {author} {\bibfnamefont {W.}~\bibnamefont {Li}}, \bibinfo {author}
  {\bibfnamefont {H.}~\bibnamefont {Masia-Roig}}, \bibinfo {author}
  {\bibfnamefont {S.}~\bibnamefont {Nix}}, \bibinfo {author} {\bibfnamefont
  {M.}~\bibnamefont {Padniuk}}, \bibinfo {author} {\bibfnamefont
  {C.}~\bibnamefont {Palm}}, \bibinfo {author} {\bibfnamefont {C.}~\bibnamefont
  {Pankow}}, \bibinfo {author} {\bibfnamefont {A.}~\bibnamefont {Penaflor}},
  \bibinfo {author} {\bibfnamefont {X.}~\bibnamefont {Peng}}, \bibinfo {author}
  {\bibfnamefont {S.}~\bibnamefont {Pustelny}}, \bibinfo {author}
  {\bibfnamefont {T.}~\bibnamefont {Scholtes}}, \bibinfo {author}
  {\bibfnamefont {J.}~\bibnamefont {Smiga}}, \bibinfo {author} {\bibfnamefont
  {J.}~\bibnamefont {Stalnaker}}, \bibinfo {author} {\bibfnamefont
  {A.}~\bibnamefont {Weis}}, \bibinfo {author} {\bibfnamefont {A.}~\bibnamefont
  {Wickenbrock}},\ and\ \bibinfo {author} {\bibfnamefont {D.}~\bibnamefont
  {Wurm}},\ }\bibfield  {title} {\bibinfo {title} {Characterization of the
  global network of optical magnetometers to search for exotic physics
  ({GNOME})},\ }\href {https://doi.org/10.1016/j.dark.2018.10.002} {\bibfield
  {journal} {\bibinfo  {journal} {Phys. Dark Universe}\ }\textbf {\bibinfo
  {volume} {22}},\ \bibinfo {pages} {162} (\bibinfo {year} {2018})}\BibitemShut
  {NoStop}%
\bibitem [{\citenamefont {Afach}\ \emph {et~al.}(2021)\citenamefont {Afach},
  \citenamefont {Buchler}, \citenamefont {Budker}, \citenamefont {Dailey},
  \citenamefont {Derevianko}, \citenamefont {Dumont}, \citenamefont {Figueroa},
  \citenamefont {Gerhardt}, \citenamefont {Grujić} \emph
  {et~al.}}]{afach_search_2021}%
  \BibitemOpen
  \bibfield  {author} {\bibinfo {author} {\bibfnamefont {S.}~\bibnamefont
  {Afach}}, \bibinfo {author} {\bibfnamefont {B.~C.}\ \bibnamefont {Buchler}},
  \bibinfo {author} {\bibfnamefont {D.}~\bibnamefont {Budker}}, \bibinfo
  {author} {\bibfnamefont {C.}~\bibnamefont {Dailey}}, \bibinfo {author}
  {\bibfnamefont {A.}~\bibnamefont {Derevianko}}, \bibinfo {author}
  {\bibfnamefont {V.}~\bibnamefont {Dumont}}, \bibinfo {author} {\bibfnamefont
  {N.~L.}\ \bibnamefont {Figueroa}}, \bibinfo {author} {\bibfnamefont
  {I.}~\bibnamefont {Gerhardt}}, \bibinfo {author} {\bibfnamefont {Z.~D.}\
  \bibnamefont {Grujić}}, \emph {et~al.},\ }\bibfield  {title} {\bibinfo
  {title} {Search for topological defect dark matter with a global network of
  optical magnetometers},\ }\href {https://doi.org/10.1038/s41567-021-01393-y}
  {\bibfield  {journal} {\bibinfo  {journal} {Nature Physics}\ }\textbf
  {\bibinfo {volume} {17}},\ \bibinfo {pages} {1396} (\bibinfo {year}
  {2021})}\BibitemShut {NoStop}%
\bibitem [{\citenamefont {Jackson~Kimball}\ \emph {et~al.}(2023)\citenamefont
  {Jackson~Kimball}, \citenamefont {Budker}, \citenamefont {Chupp},
  \citenamefont {Geraci}, \citenamefont {Kolkowitz}, \citenamefont {Singh},\
  and\ \citenamefont {Sushkov}}]{kimball2023probing}%
  \BibitemOpen
  \bibfield  {author} {\bibinfo {author} {\bibfnamefont {D.~F.}\ \bibnamefont
  {Jackson~Kimball}}, \bibinfo {author} {\bibfnamefont {D.}~\bibnamefont
  {Budker}}, \bibinfo {author} {\bibfnamefont {T.~E.}\ \bibnamefont {Chupp}},
  \bibinfo {author} {\bibfnamefont {A.~A.}\ \bibnamefont {Geraci}}, \bibinfo
  {author} {\bibfnamefont {S.}~\bibnamefont {Kolkowitz}}, \bibinfo {author}
  {\bibfnamefont {J.~T.}\ \bibnamefont {Singh}},\ and\ \bibinfo {author}
  {\bibfnamefont {A.~O.}\ \bibnamefont {Sushkov}},\ }\bibfield  {title}
  {\bibinfo {title} {Probing fundamental physics with spin-based quantum
  sensors},\ }\href {https://doi.org/10.1103/PhysRevA.108.010101} {\bibfield
  {journal} {\bibinfo  {journal} {Phys. Rev. A}\ }\textbf {\bibinfo {volume}
  {108}},\ \bibinfo {pages} {010101} (\bibinfo {year} {2023})}\BibitemShut
  {NoStop}%
\bibitem [{\citenamefont {Budker}\ and\ \citenamefont
  {Romalis}(2007)}]{budker_optical_2007}%
  \BibitemOpen
  \bibfield  {author} {\bibinfo {author} {\bibfnamefont {D.}~\bibnamefont
  {Budker}}\ and\ \bibinfo {author} {\bibfnamefont {M.}~\bibnamefont
  {Romalis}},\ }\bibfield  {title} {\bibinfo {title} {Optical magnetometry},\
  }\href {https://doi.org/10.1038/nphys566} {\bibfield  {journal} {\bibinfo
  {journal} {Nature Physics}\ }\textbf {\bibinfo {volume} {3}},\ \bibinfo
  {pages} {227} (\bibinfo {year} {2007})}\BibitemShut {NoStop}%
\bibitem [{\citenamefont {Budker}\ and\ \citenamefont
  {Jackson~Kimball}(2013)}]{budker_optical_2013}%
  \BibitemOpen
  \bibinfo {editor} {\bibfnamefont {D.}~\bibnamefont {Budker}}\ and\ \bibinfo
  {editor} {\bibfnamefont {D.~F.}\ \bibnamefont {Jackson~Kimball}},\ eds.,\
  \href
  {https://www.cambridge.org/core/books/optical-magnetometry/33E638960A4E6B159DF83F82BBDDC5E1}
  {\emph {\bibinfo {title} {Optical magnetometry}}}\ (\bibinfo  {publisher}
  {Cambridge University Press},\ \bibinfo {year} {2013})\BibitemShut {NoStop}%
\bibitem [{\citenamefont {Pospelov}\ \emph {et~al.}(2013)\citenamefont
  {Pospelov}, \citenamefont {Pustelny}, \citenamefont {Ledbetter},
  \citenamefont {Jackson~Kimball}, \citenamefont {Gawlik},\ and\ \citenamefont
  {Budker}}]{Pos13}%
  \BibitemOpen
  \bibfield  {author} {\bibinfo {author} {\bibfnamefont {M.}~\bibnamefont
  {Pospelov}}, \bibinfo {author} {\bibfnamefont {S.}~\bibnamefont {Pustelny}},
  \bibinfo {author} {\bibfnamefont {M.~P.}\ \bibnamefont {Ledbetter}}, \bibinfo
  {author} {\bibfnamefont {D.~F.}\ \bibnamefont {Jackson~Kimball}}, \bibinfo
  {author} {\bibfnamefont {W.}~\bibnamefont {Gawlik}},\ and\ \bibinfo {author}
  {\bibfnamefont {D.}~\bibnamefont {Budker}},\ }\bibfield  {title} {\bibinfo
  {title} {{Detecting Domain Walls of Axionlike Models Using Terrestrial
  Experiments}},\ }\href {https://doi.org/10.1103/PhysRevLett.110.021803}
  {\bibfield  {journal} {\bibinfo  {journal} {Phys. Rev. Lett.}\ }\textbf
  {\bibinfo {volume} {110}},\ \bibinfo {pages} {021803} (\bibinfo {year}
  {2013})}\BibitemShut {NoStop}%
\bibitem [{\citenamefont {Masia-Roig}\ \emph {et~al.}(2020)\citenamefont
  {Masia-Roig}, \citenamefont {Smiga}, \citenamefont {Budker}, \citenamefont
  {Dumont}, \citenamefont {Grujic}, \citenamefont {Kim}, \citenamefont
  {Jackson~Kimball}, \citenamefont {Lebedev}, \citenamefont {Monroy},
  \citenamefont {Pustelny} \emph {et~al.}}]{Mas20}%
  \BibitemOpen
  \bibfield  {author} {\bibinfo {author} {\bibfnamefont {H.}~\bibnamefont
  {Masia-Roig}}, \bibinfo {author} {\bibfnamefont {J.~A.}\ \bibnamefont
  {Smiga}}, \bibinfo {author} {\bibfnamefont {D.}~\bibnamefont {Budker}},
  \bibinfo {author} {\bibfnamefont {V.}~\bibnamefont {Dumont}}, \bibinfo
  {author} {\bibfnamefont {Z.}~\bibnamefont {Grujic}}, \bibinfo {author}
  {\bibfnamefont {D.}~\bibnamefont {Kim}}, \bibinfo {author} {\bibfnamefont
  {D.~F.}\ \bibnamefont {Jackson~Kimball}}, \bibinfo {author} {\bibfnamefont
  {V.}~\bibnamefont {Lebedev}}, \bibinfo {author} {\bibfnamefont
  {M.}~\bibnamefont {Monroy}}, \bibinfo {author} {\bibfnamefont
  {S.}~\bibnamefont {Pustelny}}, \emph {et~al.},\ }\bibfield  {title} {\bibinfo
  {title} {Analysis method for detecting topological defect dark matter with a
  global magnetometer network},\ }\href
  {https://doi.org/10.1016/j.dark.2020.100494} {\bibfield  {journal} {\bibinfo
  {journal} {Phys. Dark Universe}\ }\textbf {\bibinfo {volume} {28}},\ \bibinfo
  {pages} {100494} (\bibinfo {year} {2020})}\BibitemShut {NoStop}%
\bibitem [{\citenamefont {Kim}\ \emph {et~al.}(2022)\citenamefont {Kim},
  \citenamefont {Jackson~Kimball}, \citenamefont {Masia-Roig}, \citenamefont
  {Smiga}, \citenamefont {Wickenbrock}, \citenamefont {Budker}, \citenamefont
  {Kim}, \citenamefont {Shin},\ and\ \citenamefont
  {Semertzidis}}]{kim2022machine}%
  \BibitemOpen
  \bibfield  {author} {\bibinfo {author} {\bibfnamefont {D.}~\bibnamefont
  {Kim}}, \bibinfo {author} {\bibfnamefont {D.~F.}\ \bibnamefont
  {Jackson~Kimball}}, \bibinfo {author} {\bibfnamefont {H.}~\bibnamefont
  {Masia-Roig}}, \bibinfo {author} {\bibfnamefont {J.~A.}\ \bibnamefont
  {Smiga}}, \bibinfo {author} {\bibfnamefont {A.}~\bibnamefont {Wickenbrock}},
  \bibinfo {author} {\bibfnamefont {D.}~\bibnamefont {Budker}}, \bibinfo
  {author} {\bibfnamefont {Y.}~\bibnamefont {Kim}}, \bibinfo {author}
  {\bibfnamefont {Y.~C.}\ \bibnamefont {Shin}},\ and\ \bibinfo {author}
  {\bibfnamefont {Y.~K.}\ \bibnamefont {Semertzidis}},\ }\bibfield  {title}
  {\bibinfo {title} {A machine learning algorithm for direct detection of
  axion-like particle domain walls},\ }\href
  {https://doi.org/10.1016/j.dark.2022.101118} {\bibfield  {journal} {\bibinfo
  {journal} {Phys. Dark Universe}\ }\textbf {\bibinfo {volume} {37}},\ \bibinfo
  {pages} {101118} (\bibinfo {year} {2022})}\BibitemShut {NoStop}%
\bibitem [{\citenamefont {Jackson~Kimball}\ \emph {et~al.}(2018)\citenamefont
  {Jackson~Kimball}, \citenamefont {Budker}, \citenamefont {Eby}, \citenamefont
  {Pospelov}, \citenamefont {Pustelny}, \citenamefont {Scholtes}, \citenamefont
  {Stadnik}, \citenamefont {Weis},\ and\ \citenamefont
  {Wickenbrock}}]{Kim18AxionStars}%
  \BibitemOpen
  \bibfield  {author} {\bibinfo {author} {\bibfnamefont {D.~F.}\ \bibnamefont
  {Jackson~Kimball}}, \bibinfo {author} {\bibfnamefont {D.}~\bibnamefont
  {Budker}}, \bibinfo {author} {\bibfnamefont {J.}~\bibnamefont {Eby}},
  \bibinfo {author} {\bibfnamefont {M.}~\bibnamefont {Pospelov}}, \bibinfo
  {author} {\bibfnamefont {S.}~\bibnamefont {Pustelny}}, \bibinfo {author}
  {\bibfnamefont {T.}~\bibnamefont {Scholtes}}, \bibinfo {author}
  {\bibfnamefont {Y.~V.}\ \bibnamefont {Stadnik}}, \bibinfo {author}
  {\bibfnamefont {A.}~\bibnamefont {Weis}},\ and\ \bibinfo {author}
  {\bibfnamefont {A.}~\bibnamefont {Wickenbrock}},\ }\bibfield  {title}
  {\bibinfo {title} {{Searching for axion stars and Q-balls with a terrestrial
  magnetometer network}},\ }\href {https://doi.org/10.1103/PhysRevD.97.043002}
  {\bibfield  {journal} {\bibinfo  {journal} {Phys. Rev. D}\ }\textbf {\bibinfo
  {volume} {97}},\ \bibinfo {pages} {043002} (\bibinfo {year}
  {2018})}\BibitemShut {NoStop}%
\bibitem [{\citenamefont {Banerjee}\ \emph
  {et~al.}(2020{\natexlab{a}})\citenamefont {Banerjee}, \citenamefont {Budker},
  \citenamefont {Eby}, \citenamefont {Kim},\ and\ \citenamefont
  {Perez}}]{banerjee2020relaxion}%
  \BibitemOpen
  \bibfield  {author} {\bibinfo {author} {\bibfnamefont {A.}~\bibnamefont
  {Banerjee}}, \bibinfo {author} {\bibfnamefont {D.}~\bibnamefont {Budker}},
  \bibinfo {author} {\bibfnamefont {J.}~\bibnamefont {Eby}}, \bibinfo {author}
  {\bibfnamefont {H.}~\bibnamefont {Kim}},\ and\ \bibinfo {author}
  {\bibfnamefont {G.}~\bibnamefont {Perez}},\ }\bibfield  {title} {\bibinfo
  {title} {Relaxion stars and their detection via atomic physics},\ }\href
  {https://doi.org/10.1038/s42005-019-0260-3} {\bibfield  {journal} {\bibinfo
  {journal} {Commun. Phys.}\ }\textbf {\bibinfo {volume} {3}},\ \bibinfo
  {pages} {1} (\bibinfo {year} {2020}{\natexlab{a}})}\BibitemShut {NoStop}%
\bibitem [{\citenamefont {Banerjee}\ \emph
  {et~al.}(2020{\natexlab{b}})\citenamefont {Banerjee}, \citenamefont {Budker},
  \citenamefont {Eby}, \citenamefont {Flambaum}, \citenamefont {Kim},
  \citenamefont {Matsedonskyi},\ and\ \citenamefont
  {Perez}}]{banerjee2020searching}%
  \BibitemOpen
  \bibfield  {author} {\bibinfo {author} {\bibfnamefont {A.}~\bibnamefont
  {Banerjee}}, \bibinfo {author} {\bibfnamefont {D.}~\bibnamefont {Budker}},
  \bibinfo {author} {\bibfnamefont {J.}~\bibnamefont {Eby}}, \bibinfo {author}
  {\bibfnamefont {V.~V.}\ \bibnamefont {Flambaum}}, \bibinfo {author}
  {\bibfnamefont {H.}~\bibnamefont {Kim}}, \bibinfo {author} {\bibfnamefont
  {O.}~\bibnamefont {Matsedonskyi}},\ and\ \bibinfo {author} {\bibfnamefont
  {G.}~\bibnamefont {Perez}},\ }\bibfield  {title} {\bibinfo {title} {Searching
  for earth/solar axion halos},\ }\href
  {https://doi.org/10.1007/JHEP09(2020)004} {\bibfield  {journal} {\bibinfo
  {journal} {J. High Energy Phys}\ }\textbf {\bibinfo {volume} {2020}},\
  \bibinfo {pages} {4} (\bibinfo {year} {2020}{\natexlab{b}})}\BibitemShut
  {NoStop}%
\bibitem [{\citenamefont {Budker}\ \emph {et~al.}(2023)\citenamefont {Budker},
  \citenamefont {Eby}, \citenamefont {Gorghetto}, \citenamefont {Jiang},\ and\
  \citenamefont {Perez}}]{budker2023generic}%
  \BibitemOpen
  \bibfield  {author} {\bibinfo {author} {\bibfnamefont {D.}~\bibnamefont
  {Budker}}, \bibinfo {author} {\bibfnamefont {J.}~\bibnamefont {Eby}},
  \bibinfo {author} {\bibfnamefont {M.}~\bibnamefont {Gorghetto}}, \bibinfo
  {author} {\bibfnamefont {M.}~\bibnamefont {Jiang}},\ and\ \bibinfo {author}
  {\bibfnamefont {G.}~\bibnamefont {Perez}},\ }\href@noop {} {\bibinfo {title}
  {A generic formation mechanism of ultralight dark matter solar halos}}
  (\bibinfo {year} {2023}),\ \Eprint {https://arxiv.org/abs/2306.12477}
  {arXiv:2306.12477} \BibitemShut {NoStop}%
\bibitem [{\citenamefont {Dailey}\ \emph {et~al.}(2021)\citenamefont {Dailey},
  \citenamefont {Bradley}, \citenamefont {Jackson~Kimball}, \citenamefont
  {Sulai}, \citenamefont {Pustelny}, \citenamefont {Wickenbrock},\ and\
  \citenamefont {Derevianko}}]{dailey2021quantum}%
  \BibitemOpen
  \bibfield  {author} {\bibinfo {author} {\bibfnamefont {C.}~\bibnamefont
  {Dailey}}, \bibinfo {author} {\bibfnamefont {C.}~\bibnamefont {Bradley}},
  \bibinfo {author} {\bibfnamefont {D.~F.}\ \bibnamefont {Jackson~Kimball}},
  \bibinfo {author} {\bibfnamefont {I.~A.}\ \bibnamefont {Sulai}}, \bibinfo
  {author} {\bibfnamefont {S.}~\bibnamefont {Pustelny}}, \bibinfo {author}
  {\bibfnamefont {A.}~\bibnamefont {Wickenbrock}},\ and\ \bibinfo {author}
  {\bibfnamefont {A.}~\bibnamefont {Derevianko}},\ }\bibfield  {title}
  {\bibinfo {title} {Quantum sensor networks as exotic field telescopes for
  multi-messenger astronomy},\ }\href
  {https://doi.org/10.1038/s41550-020-01242-7} {\bibfield  {journal} {\bibinfo
  {journal} {Nat. Astron.}\ }\textbf {\bibinfo {volume} {5}},\ \bibinfo {pages}
  {150} (\bibinfo {year} {2021})}\BibitemShut {NoStop}%
\bibitem [{\citenamefont {Masia-Roig}\ \emph {et~al.}(2023)\citenamefont
  {Masia-Roig}, \citenamefont {Figueroa}, \citenamefont {Bordon}, \citenamefont
  {Smiga}, \citenamefont {Stadnik}, \citenamefont {Budker}, \citenamefont
  {Centers}, \citenamefont {Gramolin}, \citenamefont {Hamilton}, \citenamefont
  {Khamis} \emph {et~al.}}]{masia2023intensity}%
  \BibitemOpen
  \bibfield  {author} {\bibinfo {author} {\bibfnamefont {H.}~\bibnamefont
  {Masia-Roig}}, \bibinfo {author} {\bibfnamefont {N.~L.}\ \bibnamefont
  {Figueroa}}, \bibinfo {author} {\bibfnamefont {A.}~\bibnamefont {Bordon}},
  \bibinfo {author} {\bibfnamefont {J.~A.}\ \bibnamefont {Smiga}}, \bibinfo
  {author} {\bibfnamefont {Y.~V.}\ \bibnamefont {Stadnik}}, \bibinfo {author}
  {\bibfnamefont {D.}~\bibnamefont {Budker}}, \bibinfo {author} {\bibfnamefont
  {G.~P.}\ \bibnamefont {Centers}}, \bibinfo {author} {\bibfnamefont {A.~V.}\
  \bibnamefont {Gramolin}}, \bibinfo {author} {\bibfnamefont {P.~S.}\
  \bibnamefont {Hamilton}}, \bibinfo {author} {\bibfnamefont {S.}~\bibnamefont
  {Khamis}}, \emph {et~al.},\ }\bibfield  {title} {\bibinfo {title} {Intensity
  interferometry for ultralight bosonic dark matter detection},\ }\href
  {https://doi.org/10.1103/PhysRevD.108.015003} {\bibfield  {journal} {\bibinfo
   {journal} {Phys. Rev. D}\ }\textbf {\bibinfo {volume} {108}},\ \bibinfo
  {pages} {015003} (\bibinfo {year} {2023})}\BibitemShut {NoStop}%
\bibitem [{\citenamefont {Padniuk}\ \emph
  {et~al.}(2022{\natexlab{a}})\citenamefont {Padniuk}, \citenamefont
  {Kopciuch}, \citenamefont {Cipolletti}, \citenamefont {Wickenbrock},
  \citenamefont {Budker},\ and\ \citenamefont
  {Pustelny}}]{padniuk2022response}%
  \BibitemOpen
  \bibfield  {author} {\bibinfo {author} {\bibfnamefont {M.}~\bibnamefont
  {Padniuk}}, \bibinfo {author} {\bibfnamefont {M.}~\bibnamefont {Kopciuch}},
  \bibinfo {author} {\bibfnamefont {R.}~\bibnamefont {Cipolletti}}, \bibinfo
  {author} {\bibfnamefont {A.}~\bibnamefont {Wickenbrock}}, \bibinfo {author}
  {\bibfnamefont {D.}~\bibnamefont {Budker}},\ and\ \bibinfo {author}
  {\bibfnamefont {S.}~\bibnamefont {Pustelny}},\ }\bibfield  {title} {\bibinfo
  {title} {Response of atomic spin-based sensors to magnetic and nonmagnetic
  perturbations},\ }\href {https://doi.org/10.1038/s41598-021-03609-w}
  {\bibfield  {journal} {\bibinfo  {journal} {Sci. Rep.}\ }\textbf {\bibinfo
  {volume} {12}},\ \bibinfo {pages} {324} (\bibinfo {year}
  {2022}{\natexlab{a}})}\BibitemShut {NoStop}%
\bibitem [{\citenamefont {Kornack}\ and\ \citenamefont
  {Romalis}(2002{\natexlab{a}})}]{kornack2002dynamics}%
  \BibitemOpen
  \bibfield  {author} {\bibinfo {author} {\bibfnamefont {T.~W.}\ \bibnamefont
  {Kornack}}\ and\ \bibinfo {author} {\bibfnamefont {M.~V.}\ \bibnamefont
  {Romalis}},\ }\bibfield  {title} {\bibinfo {title} {Dynamics of two
  overlapping spin ensembles interacting by spin exchange},\ }\href
  {https://doi.org/10.1103/PhysRevLett.89.253002} {\bibfield  {journal}
  {\bibinfo  {journal} {Phys. Rev. Lett.}\ }\textbf {\bibinfo {volume} {89}},\
  \bibinfo {pages} {253002} (\bibinfo {year} {2002}{\natexlab{a}})}\BibitemShut
  {NoStop}%
\bibitem [{\citenamefont {Kornack}()}]{kornack_test_nodate}%
  \BibitemOpen
  \bibfield  {author} {\bibinfo {author} {\bibfnamefont {T.~W.}\ \bibnamefont
  {Kornack}},\ }\emph {\bibinfo {title} {A test of {CPT} and {Lorentz} symmetry
  using a potassium-helium-3 co-magnetometer}},\ \href
  {https://www.proquest.com/docview/305415550/abstract/20EE3BBAD3334C42PQ/1}
  {\bibinfo {type} {Ph.{D}.}},\ \bibinfo  {school} {Princeton University},
  \bibinfo {address} {United States -- New Jersey}\BibitemShut {NoStop}%
\bibitem [{\citenamefont {Vasilakis}()}]{vasilakis_precision_nodate}%
  \BibitemOpen
  \bibfield  {author} {\bibinfo {author} {\bibfnamefont {G.}~\bibnamefont
  {Vasilakis}},\ }\emph {\bibinfo {title} {Precision measurements of spin
  interactions with high density atomic vapors}},\ \href
  {https://www.proquest.com/docview/898956586/abstract/BF95BCE6F18E4C11PQ/1}
  {\bibinfo {type} {Ph.{D}.}},\ \bibinfo  {school} {Princeton University},
  \bibinfo {address} {United States -- New Jersey}\BibitemShut {NoStop}%
\bibitem [{\citenamefont {Fan}\ \emph {et~al.}(2019)\citenamefont {Fan},
  \citenamefont {Quan}, \citenamefont {Zhang}, \citenamefont {Xing},\ and\
  \citenamefont {Liu}}]{FanIEEE2019}%
  \BibitemOpen
  \bibfield  {author} {\bibinfo {author} {\bibfnamefont {W.}~\bibnamefont
  {Fan}}, \bibinfo {author} {\bibfnamefont {W.}~\bibnamefont {Quan}}, \bibinfo
  {author} {\bibfnamefont {W.}~\bibnamefont {Zhang}}, \bibinfo {author}
  {\bibfnamefont {L.}~\bibnamefont {Xing}},\ and\ \bibinfo {author}
  {\bibfnamefont {G.}~\bibnamefont {Liu}},\ }\bibfield  {title} {\bibinfo
  {title} {Analysis on the magnetic field response for nuclear spin
  co-magnetometer operated in spin-exchange relaxation-free regime},\ }\href
  {https://doi.org/10.1109/ACCESS.2019.2902181} {\bibfield  {journal} {\bibinfo
   {journal} {IEEE Access}\ }\textbf {\bibinfo {volume} {7}},\ \bibinfo {pages}
  {28574} (\bibinfo {year} {2019})}\BibitemShut {NoStop}%
\bibitem [{\citenamefont {Kornack}\ \emph
  {et~al.}(2005{\natexlab{b}})\citenamefont {Kornack}, \citenamefont {Ghosh},\
  and\ \citenamefont {Romalis}}]{kornack_nuclear_2005}%
  \BibitemOpen
  \bibfield  {author} {\bibinfo {author} {\bibfnamefont {T.~W.}\ \bibnamefont
  {Kornack}}, \bibinfo {author} {\bibfnamefont {R.~K.}\ \bibnamefont {Ghosh}},\
  and\ \bibinfo {author} {\bibfnamefont {M.~V.}\ \bibnamefont {Romalis}},\
  }\bibfield  {title} {\bibinfo {title} {Nuclear {Spin} {Gyroscope} {Based} on
  an {Atomic} {Comagnetometer}},\ }\href
  {https://doi.org/10.1103/PhysRevLett.95.230801} {\bibfield  {journal}
  {\bibinfo  {journal} {Phys. Rev. Lett.}\ }\textbf {\bibinfo {volume} {95}},\
  \bibinfo {pages} {230801} (\bibinfo {year} {2005}{\natexlab{b}})}\BibitemShut
  {NoStop}%
\bibitem [{\citenamefont {Padniuk}\ \emph
  {et~al.}(2022{\natexlab{b}})\citenamefont {Padniuk}, \citenamefont
  {Kopciuch}, \citenamefont {Cipolletti}, \citenamefont {Wickenbrock},
  \citenamefont {Budker},\ and\ \citenamefont
  {Pustelny}}]{padniuk_response_2022}%
  \BibitemOpen
  \bibfield  {author} {\bibinfo {author} {\bibfnamefont {M.}~\bibnamefont
  {Padniuk}}, \bibinfo {author} {\bibfnamefont {M.}~\bibnamefont {Kopciuch}},
  \bibinfo {author} {\bibfnamefont {R.}~\bibnamefont {Cipolletti}}, \bibinfo
  {author} {\bibfnamefont {A.}~\bibnamefont {Wickenbrock}}, \bibinfo {author}
  {\bibfnamefont {D.}~\bibnamefont {Budker}},\ and\ \bibinfo {author}
  {\bibfnamefont {S.}~\bibnamefont {Pustelny}},\ }\bibfield  {title} {\bibinfo
  {title} {Response of atomic spin-based sensors to magnetic and nonmagnetic
  perturbations},\ }\href {https://doi.org/10.1038/s41598-021-03609-w}
  {\bibfield  {journal} {\bibinfo  {journal} {Sci. Rep.}\ }\textbf {\bibinfo
  {volume} {12}},\ \bibinfo {pages} {324} (\bibinfo {year}
  {2022}{\natexlab{b}})}\BibitemShut {NoStop}%
\bibitem [{\citenamefont {Hasegawa}(1959)}]{Hasegawa1959}%
  \BibitemOpen
  \bibfield  {author} {\bibinfo {author} {\bibfnamefont {H.}~\bibnamefont
  {Hasegawa}},\ }\bibfield  {title} {\bibinfo {title} {{Dynamical Properties of
  $s-d$ Interaction}},\ }\href {https://doi.org/10.1143/PTP.21.483} {\bibfield
  {journal} {\bibinfo  {journal} {Progress of Theoretical Physics}\ }\textbf
  {\bibinfo {volume} {21}},\ \bibinfo {pages} {483} (\bibinfo {year}
  {1959})}\BibitemShut {NoStop}%
\bibitem [{\citenamefont {Kornack}\ and\ \citenamefont
  {Romalis}(2002{\natexlab{b}})}]{kornack_dynamics_2002}%
  \BibitemOpen
  \bibfield  {author} {\bibinfo {author} {\bibfnamefont {T.~W.}\ \bibnamefont
  {Kornack}}\ and\ \bibinfo {author} {\bibfnamefont {M.~V.}\ \bibnamefont
  {Romalis}},\ }\bibfield  {title} {\bibinfo {title} {Dynamics of {Two}
  {Overlapping} {Spin} {Ensembles} {Interacting} by {Spin} {Exchange}},\ }\href
  {https://doi.org/10.1103/PhysRevLett.89.253002} {\bibfield  {journal}
  {\bibinfo  {journal} {Phys. Rev. Lett.}\ }\textbf {\bibinfo {volume} {89}},\
  \bibinfo {pages} {253002} (\bibinfo {year} {2002}{\natexlab{b}})}\BibitemShut
  {NoStop}%
\bibitem [{\citenamefont {Allred}\ \emph {et~al.}(2002)\citenamefont {Allred},
  \citenamefont {Lyman}, \citenamefont {Kornack},\ and\ \citenamefont
  {Romalis}}]{Allred2002}%
  \BibitemOpen
  \bibfield  {author} {\bibinfo {author} {\bibfnamefont {J.~C.}\ \bibnamefont
  {Allred}}, \bibinfo {author} {\bibfnamefont {R.~N.}\ \bibnamefont {Lyman}},
  \bibinfo {author} {\bibfnamefont {T.~W.}\ \bibnamefont {Kornack}},\ and\
  \bibinfo {author} {\bibfnamefont {M.~V.}\ \bibnamefont {Romalis}},\
  }\bibfield  {title} {\bibinfo {title} {High-sensitivity atomic magnetometer
  unaffected by spin-exchange relaxation},\ }\href
  {https://doi.org/10.1103/PhysRevLett.89.130801} {\bibfield  {journal}
  {\bibinfo  {journal} {Phys. Rev. Lett.}\ }\textbf {\bibinfo {volume} {89}},\
  \bibinfo {pages} {130801} (\bibinfo {year} {2002})}\BibitemShut {NoStop}%
\bibitem [{\citenamefont {Ben-Amar~Baranga}\ \emph {et~al.}(1998)\citenamefont
  {Ben-Amar~Baranga}, \citenamefont {Appelt}, \citenamefont {Romalis},
  \citenamefont {Erickson}, \citenamefont {Young}, \citenamefont {Cates},\ and\
  \citenamefont {Happer}}]{baranga1998}%
  \BibitemOpen
  \bibfield  {author} {\bibinfo {author} {\bibfnamefont {A.}~\bibnamefont
  {Ben-Amar~Baranga}}, \bibinfo {author} {\bibfnamefont {S.}~\bibnamefont
  {Appelt}}, \bibinfo {author} {\bibfnamefont {M.~V.}\ \bibnamefont {Romalis}},
  \bibinfo {author} {\bibfnamefont {C.~J.}\ \bibnamefont {Erickson}}, \bibinfo
  {author} {\bibfnamefont {A.~R.}\ \bibnamefont {Young}}, \bibinfo {author}
  {\bibfnamefont {G.~D.}\ \bibnamefont {Cates}},\ and\ \bibinfo {author}
  {\bibfnamefont {W.}~\bibnamefont {Happer}},\ }\bibfield  {title} {\bibinfo
  {title} {Polarization of ${}^{3}\mathrm{He}$ by spin exchange with optically
  pumped {Rb} and {K} vapors},\ }\href
  {https://doi.org/10.1103/PhysRevLett.80.2801} {\bibfield  {journal} {\bibinfo
   {journal} {Phys. Rev. Lett.}\ }\textbf {\bibinfo {volume} {80}},\ \bibinfo
  {pages} {2801} (\bibinfo {year} {1998})}\BibitemShut {NoStop}%
\bibitem [{\citenamefont {Kimball}(2015)}]{kimball_nuclear_2015}%
  \BibitemOpen
  \bibfield  {author} {\bibinfo {author} {\bibfnamefont {D.~F.~J.}\
  \bibnamefont {Kimball}},\ }\bibfield  {title} {\bibinfo {title} {Nuclear spin
  content and constraints on exotic spin-dependent couplings},\ }\href
  {https://doi.org/10.1088/1367-2630/17/7/073008} {\bibfield  {journal}
  {\bibinfo  {journal} {New J. Phys.}\ }\textbf {\bibinfo {volume} {17}},\
  \bibinfo {pages} {073008} (\bibinfo {year} {2015})}\BibitemShut {NoStop}%
\bibitem [{\citenamefont {Klinger}\ \emph {et~al.}(2023)\citenamefont
  {Klinger}, \citenamefont {Liu}, \citenamefont {Padniuk}, \citenamefont
  {Engler}, \citenamefont {Kornack}, \citenamefont {Pustelny}, \citenamefont
  {Jackson~Kimball}, \citenamefont {Budker},\ and\ \citenamefont
  {Wickenbrock}}]{klingerPRA2023}%
  \BibitemOpen
  \bibfield  {author} {\bibinfo {author} {\bibfnamefont {E.}~\bibnamefont
  {Klinger}}, \bibinfo {author} {\bibfnamefont {T.}~\bibnamefont {Liu}},
  \bibinfo {author} {\bibfnamefont {M.}~\bibnamefont {Padniuk}}, \bibinfo
  {author} {\bibfnamefont {M.}~\bibnamefont {Engler}}, \bibinfo {author}
  {\bibfnamefont {T.}~\bibnamefont {Kornack}}, \bibinfo {author} {\bibfnamefont
  {S.}~\bibnamefont {Pustelny}}, \bibinfo {author} {\bibfnamefont {D.~F.}\
  \bibnamefont {Jackson~Kimball}}, \bibinfo {author} {\bibfnamefont
  {D.}~\bibnamefont {Budker}},\ and\ \bibinfo {author} {\bibfnamefont
  {A.}~\bibnamefont {Wickenbrock}},\ }\bibfield  {title} {\bibinfo {title}
  {Optimization of nuclear polarization in an alkali-noble gas
  comagnetometer},\ }\href {https://doi.org/10.1103/PhysRevApplied.19.044092}
  {\bibfield  {journal} {\bibinfo  {journal} {Phys. Rev. Appl.}\ }\textbf
  {\bibinfo {volume} {19}},\ \bibinfo {pages} {044092} (\bibinfo {year}
  {2023})}\BibitemShut {NoStop}%
\bibitem [{\citenamefont {Babcock}\ \emph {et~al.}(2003)\citenamefont
  {Babcock}, \citenamefont {Nelson}, \citenamefont {Kadlecek}, \citenamefont
  {Driehuys}, \citenamefont {Anderson}, \citenamefont {Hersman},\ and\
  \citenamefont {Walker}}]{babcock_hybrid_2003}%
  \BibitemOpen
  \bibfield  {author} {\bibinfo {author} {\bibfnamefont {E.}~\bibnamefont
  {Babcock}}, \bibinfo {author} {\bibfnamefont {I.}~\bibnamefont {Nelson}},
  \bibinfo {author} {\bibfnamefont {S.}~\bibnamefont {Kadlecek}}, \bibinfo
  {author} {\bibfnamefont {B.}~\bibnamefont {Driehuys}}, \bibinfo {author}
  {\bibfnamefont {L.~W.}\ \bibnamefont {Anderson}}, \bibinfo {author}
  {\bibfnamefont {F.~W.}\ \bibnamefont {Hersman}},\ and\ \bibinfo {author}
  {\bibfnamefont {T.~G.}\ \bibnamefont {Walker}},\ }\bibfield  {title}
  {\bibinfo {title} {Hybrid {Spin}-{Exchange} {Optical} {Pumping} of
  $^3${He}},\ }\href {https://doi.org/10.1103/PhysRevLett.91.123003} {\bibfield
   {journal} {\bibinfo  {journal} {Phys. Rev. Lett.}\ }\textbf {\bibinfo
  {volume} {91}},\ \bibinfo {pages} {123003} (\bibinfo {year}
  {2003})}\BibitemShut {NoStop}%
\bibitem [{\citenamefont {Lee}\ \emph {et~al.}(2021)\citenamefont {Lee},
  \citenamefont {Lee},\ and\ \citenamefont {Yim}}]{Lee:21}%
  \BibitemOpen
  \bibfield  {author} {\bibinfo {author} {\bibfnamefont {D.-Y.}\ \bibnamefont
  {Lee}}, \bibinfo {author} {\bibfnamefont {S.}~\bibnamefont {Lee}},\ and\
  \bibinfo {author} {\bibfnamefont {S.~H.}\ \bibnamefont {Yim}},\ }\bibfield
  {title} {\bibinfo {title} {Measurement of a $^{129}${Xe} transverse
  relaxation rate without the influence of {Rb} polarization-induced magnetic
  gradient},\ }\href {https://doi.org/10.1364/AO.427613} {\bibfield  {journal}
  {\bibinfo  {journal} {Appl. Opt.}\ }\textbf {\bibinfo {volume} {60}},\
  \bibinfo {pages} {7290} (\bibinfo {year} {2021})}\BibitemShut {NoStop}%
\bibitem [{\citenamefont {Li}\ \emph {et~al.}(2006)\citenamefont {Li},
  \citenamefont {Wakai},\ and\ \citenamefont {Walker}}]{zhimin2006parametric}%
  \BibitemOpen
  \bibfield  {author} {\bibinfo {author} {\bibfnamefont {Z.}~\bibnamefont
  {Li}}, \bibinfo {author} {\bibfnamefont {R.~T.}\ \bibnamefont {Wakai}},\ and\
  \bibinfo {author} {\bibfnamefont {T.~G.}\ \bibnamefont {Walker}},\ }\bibfield
   {title} {\bibinfo {title} {Parametric modulation of an atomic
  magnetometer},\ }\href {https://doi.org/10.1063/1.2357553} {\bibfield
  {journal} {\bibinfo  {journal} {Appl. Phys. Lett.}\ }\textbf {\bibinfo
  {volume} {89}},\ \bibinfo {pages} {134105} (\bibinfo {year}
  {2006})}\BibitemShut {NoStop}%
\end{thebibliography}%

\clearpage

\appendix

\end{document}